\documentclass[english,two column]{revtex4-1}
\usepackage[T1]{fontenc}
\usepackage[latin9]{inputenc}
\usepackage{amsmath}
\usepackage{amssymb}
\usepackage{graphicx}
\usepackage{babel}
\usepackage{color}
\begin{document}
\title{Dynamic structure factor of two-dimensional Fermi superfluid with
Rashba spin-orbit coupling}

\author{Huaisong Zhao$^1$}
\author{Xu Yan$^2$}
\email{qdyanx@qdu.edu.cn}
\author{Shi-Guo Peng$^3$}
\email{pengshiguo@wipm.ac.cn}
\author{Peng Zou$^1$}
\email{phy.zoupeng@gmail.com}

\affiliation{$^1$College of Physics, Qingdao University, Qingdao 266071, China}
\affiliation{$^2$Industrial Research Institute of Nonwovens and Technical Textiles, Shandong Center for Engineered Nonwovens, College of Textiles and Clothing,
Qingdao University, Qingdao 266071, China}
\affiliation{$^3$State Key Laboratory of Magnetic Resonance and Atomic and Molecular Physics, Innovation Academy for Precision Measurement Science and Technology, Chinese Academy of Sciences, Wuhan 430071, China}

\begin{abstract}
We theoretically calculate the dynamic structure factor of two-dimensional
Rashba-type spin-orbit coupled (SOC) Fermi superfluid with random
phase approximation, and analyze the main characters of dynamical
excitation shown by both density and spin dynamic structure factor
during a continuous phase transition between Bardeen-Cooper-Schrieffer
superfluid and topological superfluid. Generally we find three different
excitations, including collective phonon excitation, two-atom molecular and atomic excitations, and pair-breaking
excitations due to two-branch structure of quasi-particle spectrum. It should be emphasized
that collective phonon excitation is overlapped with a gapless $DD$
type pair-breaking excitation at the critical Zeeman field $h_{c}$,
and is imparted a finite width to phonon peak when transferred momentum $\textbf{\textit{q}}$
is around Fermi vector $k_{F}$. At a much larger transferred momentum
($\textbf{\textit{q}}=4k_{F}$), the pair-breaking excitation happens earlier than two-atom
molecular excitation, which is different from the conventional Fermi
superfluid without SOC effect.
\end{abstract}
\maketitle

\section{Introduction}

Finding and distinguishing exotic matter states are interesting and
important tasks in quantum many-body physics \citep{Bloch2008rmp,Giorgini2008rmp}.
In atomic physics, the strategy that analyzing the atomic spectrum structure shown by
all possible electronic transition between atomic energy levels has been
verified to be an effective way to distinguish chemical elements.
All spectra consist of dynamical excitations information, which
can be described by optical spectrum dynamic structure factor. In
quantum many-body physics, many particles interplay with each
other and arrive at rich equilibrium matter states. Since the realization
of spin-linear (angular) momentum coupling effect in ultracold atomic
gases \citep{Lin2011s,Wang2012s,Cheuk2012s,Chen2018prl,Zhang2019prl,Williams2013,Meng2016,Burdick2016,Huang2016,Wu2016,Wang2021},
it has been possible to investigate many exotic matter states, such
as two types of stripe phase with discrete translational or rational symmetry \citep{Ho2011b,Michael2015pra,Qu2015pra},
the topological state \citep{Jiang2011m,Chen2022prr,Han2023}, etc., in these
highly controllable systems. Naturally it is interesting to know whether
it is possible to find a universal way to identify all these matter
states with dynamical excitations information.

Dynamic structure factor, which is related to the imaginary part of response function in the momentum-energy representation, is an important many-body physical quantity and includes rich dynamics about the system in a certain matter state \citep{Pitaevskii2003book}.
Experimentally dynamic structure factor can be measured by a two-photon
Bragg scattering technique \citep{Veeravalli2008b,Hoinka2017g,Biss2022,Senaratne2022,Pagano2014}, in
which two Bragg laser beams can transfer a selected transferred momentum $\textbf{\textit{q}}$
and energy $\omega$ to perturb the system. After this Bragg perturbation, the dynamic structure factor can be obtained by measuring the centre-of-mass velocity of the system \citep{Brunello2001pra}. In a small transferred momentum
$\textbf{\textit{q}}$, some momentum-related collective modes, like Goldstone phonon \citep{Hoinka2017g,Combescot2006pra,Kettmann2017},
second sound \citep{Hu2022aapps,Li2022}, Higgs mode \citep{Pekker2015arcmp,Behrle2018,Bjerlin2016,Bruun2014,Zhao2020d,Fan2022p,Phan2023},
and Leggett excitation \citep{Leggett1966n,Zhang2017t}, can be observed.
At a large $\textbf{\textit{q}}$, the dynamics is dominated by the single-particle excitation, including not only the Cooper pair-breaking excitation, but also the ideal (or interaction-revised) atomic or molecular excitation \citep{Combescot2006m,Combescot2006pra}. All
these dynamics consist of dynamical character of the system in a specific
matter state, and may display different dynamical behaviours during phase transition. So it is interesting to study these dynamical characters
of the system in different matter states according to dynamic structure factor, and check the feasibility to identify matter state by dynamic
structure factor. In our previous work, dynamic structure factor had been found that it can  display different dynamical behaviours in a few phase transitions \citep{Zou2021d,Gao2023pra,Han2022}.

In this paper, we theoretically investigate a two-dimensional (2D)
Rashba-type spin-orbit coupled (SOC) Fermi superfluid, which can experience
a continuous phase transition between a conventional Bardeen-Cooper-Schrieffer
(BCS) superfluid to a topological superfluid by continuously varying
the Zeeman field \citep{Liu2012praR}. We numerically calculate dynamic
structure factor of this system with random phase approximation \citep{AndersonPR1958,Liu2004c},
and try to find its main dynamical excitation character during phase transition. We find the dynamic structure factor presents rich excitation signals, including collective phonon excitation, molecular or atomic excitations, and four kinds of pair-breaking excitations due to two-branch structure of quasi-particle spectrum. Among all these excitations, the collective phonon excitation requires the smallest excitation energy in both the BCS and topological superfluid. In the critical point of phase
transition, one of pair-breaking excitations becomes gapless excitation. It overlaps with the phonon excitation and imparts a finite expansion width to the phonon peak in a certain transferred momentum.
This paper is organized as follows. In the next section, we will use
the motion equation of Green's function to introduce the
microscopic model of a 2D Fermi superfluid with the Rashba SOC effect,
outline the mean-field approximation, and show how to calculate the
response function with the random phase approximation. We give results
of the dynamic structure factor of both BCS and topological superfluids
in Sec. III, and give our conclusions and outlook in Sec. IV. Some
calculation details will be listed in the Appendix.

\section{Methods}

\subsection{Model and Hamiltonian}

We consider a uniform 2D Fermi gases subject to a Rashba SOC potential $V_{{\rm soc}}=-i\lambda\left(\partial_{y}+i\partial_{x}\right)$
with strength $\lambda$ and a Zeeman field $h$. The system can be
described by a model Hamiltonian
\begin{equation}\label{eq_Ham}
H=\int d^{2}r\left[\sum_{\sigma}\mathcal{H}_{\sigma}^{S}+\mathcal{H}_{{\rm SOC}}+\mathcal{H}_{{\rm int}}\right].
\end{equation}
Here $\mathcal{H}_{\sigma}^{S}=\psi_{\sigma}^{\dagger}\left[-\nabla^{2}/2m-\mu-h\sigma_{z}\right]\psi_{\sigma}$
is the single particle Hamiltonian of spin-$\sigma$ component particles
with mass $m$ in reference to the chemical potential $\mu$, and
$\psi_{\sigma}(\psi_{\sigma}^{\dagger})$ is the annihilation (generation)
operator. $\mathcal{H}_{{\rm SOC}}=\psi_{\uparrow}^{\dagger}V_{{\rm soc}}\psi_{\downarrow}+h.c.$
is the Rashba SOC Hamiltonian, and it should be noted that the strength $\lambda$ of SOC effect is isotropic in the 2D $XY$-plane. $\mathcal{H}_{{\rm int}}=U\psi_{\uparrow}^{\dagger}\psi_{\downarrow}^{\dagger}\psi_{\downarrow}\psi_{\uparrow}$
describes the contact interaction between opposite spins, in which
the strength $U$ should be regularized via $1/U=-\sum_{\textbf{\textit{k}}}1/\left(\textbf{\textit{k}}^{2}/m+E_{a}\right)$.
$E_{a}$ is the binding energy of the two-body bound state, and is often used to demonstrate the interaction strength in 2D system \citep{Bertaina2011PRL}.  Here
and in the following we have set $\hbar=k_{B}=1$ for simple. Since
we consider a uniform system with bulk density $n_{0}$, the inverse
of Fermi wave vector $k_{F}=\sqrt{2\pi n_{0}}$ and Fermi energy $E_{F}=k_{F}^{2}/2m$
are used as length and energy units, respectively.

A standard mean-field treatment is carried out to the interaction
Hamiltonian $\mathcal{H}_{{\rm int}}$ with the usual definition of order
parameter $\Delta=-U\left\langle \psi_{\downarrow}\psi_{\uparrow}\right\rangle $.
After Fourier transformation to the mean-field model Hamiltonian,
we can obtain its expression in the momentum representation, which
reads

\begin{equation}
\begin{array}{cl}
H_{{\rm mf}} & =\underset{\textbf{\textit{k}}\sigma}{\sum}\left(\xi_{\textbf{\textit{k}}}-h\sigma_{z}\right)c_{\textbf{\textit{k}}\sigma}^{\dagger}c_{\textbf{\textit{k}}\sigma}\\
 & +\underset{\textbf{\textit{k}}}{\sum}\left[\lambda\left(k_{y}+ik_{x}\right)c_{\textbf{\textit{k}}\uparrow}^{\dagger}c_{\textbf{\textit{k}}\downarrow}+h.c.\right]\\
 & -\underset{\textbf{\textit{k}}}{\sum}\left[\Delta c_{\textbf{\textit{k}}\uparrow}^{\dagger}c_{-\textbf{\textit{k}}\downarrow}^{\dagger}+\Delta^{*}c_{-\textbf{\textit{k}}\downarrow}c_{\textbf{\textit{k}}\uparrow}\right]
\end{array}
\end{equation}
with $\xi_{\textbf{\textit{\textbf{\textit{k}}}}}=\textbf{\textit{k}}^{2}/2m-\mu$. Usually the order parameter $\Delta$
is a complex number. However U(1) symmetry is spontaneously broken
in the ground state of the system, and the value for the phase of $\Delta$ is pushed
to choose a random number. Here we just set $\Delta=\Delta^{*}$.
The exact diagonalization of mean-field Hamiltonian $H_{{\rm mf}}$
can be solved with motion equation of Green's functions. Finally we
get six independent Green's functions, which are

\begin{equation}
\begin{array}{ccc}
G_{1}\left(\textbf{\textit{k}},\omega\right) & \equiv\left\langle \left\langle c_{\textbf{\textit{k}}\uparrow}|c_{\textbf{\textit{k}}\uparrow}^{\dagger}\right\rangle \right\rangle  & =\underset{l}{\sum}\left[G_{1}\right]_{\textbf{\textit{k}}}^{l}/\left(\omega-E_{\textbf{\textit{k}}}^{l}\right),\\
G_{2}\left(\textbf{\textit{k}},\omega\right) & \equiv\left\langle \left\langle c_{\textbf{\textit{k}}\downarrow}|c_{\textbf{\textit{k}}\downarrow}^{\dagger}\right\rangle \right\rangle  & =\underset{l}{\sum}\left[G_{2}\right]_{\textbf{\textit{k}}}^{l}/\left(\omega-E_{\textbf{\textit{k}}}^{l}\right),\\
\varGamma\left(\textbf{\textit{k}},\omega\right) & \equiv\left\langle \left\langle c_{\textbf{\textit{k}}\uparrow}|c_{-\textbf{\textit{k}}\downarrow}\right\rangle \right\rangle  & =\underset{l}{\sum}\left[\varGamma\right]_{\textbf{\textit{k}}}^{l}/\left(\omega-E_{\textbf{\textit{k}}}^{l}\right),\\
S\left(\textbf{\textit{k}},\omega\right) & \equiv\left\langle \left\langle c_{\textbf{\textit{k}}\downarrow}|c_{\textbf{\textit{k}}\uparrow}^{\dagger}\right\rangle \right\rangle  & =\underset{l}{\sum}\left[S\right]_{\textbf{\textit{k}}}^{l}/\left(\omega-E_{\textbf{\textit{k}}}^{l}\right),\\
F_{1}\left(\textbf{\textit{k}},\omega\right) & \equiv\left\langle \left\langle c_{\textbf{\textit{k}}\uparrow}|c_{-\textbf{\textit{k}}\uparrow}\right\rangle \right\rangle  & =\underset{l}{\sum}\left[F_{1}\right]_{\textbf{\textit{k}}}^{l}/\left(\omega-E_{\textbf{\textit{k}}}^{l}\right),\\
F_{2}\left(\textbf{\textit{k}},\omega\right) & \equiv\left\langle \left\langle c_{\textbf{\textit{k}}\downarrow}|c_{-\textbf{\textit{k}}\downarrow}\right\rangle \right\rangle  & =\underset{l}{\sum}\left[F_{2}\right]_{\textbf{\textit{k}}}^{l}/\left(\omega-E_{\textbf{\textit{k}}}^{l}\right),
\end{array}\label{eq:GF}
\end{equation}
where $l=\pm1,\pm2$ denotes respectively all four important quasi-particle
energy spectra $E_{\textbf{\textit{k}}}^{\left(+1\right)}=-E_{\textbf{\textit{k}}}^{\left(-1\right)}=U_{\textbf{\textit{k}}}$
and $E_{\textbf{\textit{k}}}^{\left(+2\right)}=-E_{\textbf{\textit{k}}}^{\left(-2\right)}=D_{\textbf{\textit{k}}}$. $U_{\textbf{\textit{k}}}$
and $D_{\textbf{\textit{k}}}$ are respectively the up- and down-branch positive quasi-particle
spectrum

\begin{equation}
U_{\textbf{\textit{k}}}=\sqrt{E_{\textbf{\textit{k}}}^{2}+h^{2}+\textbf{\textit{k}}^{2}\lambda^{2}+2\sqrt{E_{\textbf{\textit{k}}}^{2}h^{2}+\xi_{\textbf{\textit{k}}}^{2}\textbf{\textit{k}}^{2}\lambda^{2}}},\label{eq:Uk}
\end{equation}

\begin{equation}
D_{\textbf{\textit{k}}}=\sqrt{E_{\textbf{\textit{k}}}^{2}+h^{2}+\textbf{\textit{k}}^{2}\lambda^{2}-2\sqrt{E_{\textbf{\textit{k}}}^{2}h^{2}+\xi_{\textbf{\textit{k}}}^{2}\textbf{\textit{k}}^{2}\lambda^{2}}},\label{eq:Dk}
\end{equation}
with $E_{\textbf{\textit{k}}}=\sqrt{\xi_{\textbf{\textit{k}}}^2+\Delta^{2}}$. The Green's functions $S,F_{1}$and
$F_{2}$ come from the Rashba SOC Hamiltonian, and they are the odd
functions of momentum $\textbf{\textit{k}}$, which are the even functions in the Raman SOC
case. These single-particle spectra ($U_{\textbf{\textit{k}}}$ and $D_{\textbf{\textit{k}}}$) do great
influence to the static and dynamical properties of ground state.
All expressions related to $\left[G_{1}\right]_{\textbf{\textit{k}}}^{l}$,$\left[G_{2}\right]_{\textbf{\textit{k}}}^{l}$,$\left[\Gamma\right]_{\textbf{\textit{k}}}^{l}$,$\left[S\right]_{\textbf{\textit{k}}}^{l}$,$\left[F_{1}\right]_{\textbf{\textit{k}}}^{l}$
and $\left[F_{2}\right]_{\textbf{\textit{k}}}^{l}$ will be given in the appendix.

Based on the spectrum theorem, it is also easy to get all equations
of all physical quantities with the above Green's functions. For example,
we obtain spin-up and spin-down density equations

\begin{equation}
n_{1}=\sum_{\textbf{\textit{k}}}\left\langle c_{\textbf{\textit{k}}\uparrow}^{\dagger}c_{\textbf{\textit{k}}\uparrow}\right\rangle =-\frac{1}{\pi}\sum_{\textbf{\textit{k}}}\int d\omega\frac{{\rm Im}\left[G_{1}\left(\textbf{\textit{k}},\omega\right)\right]}{e^{\omega/T}+1},
\end{equation}

\begin{equation}
n_{2}=\sum_{\textbf{\textit{k}}}\left\langle c_{\textbf{\textit{k}}\downarrow}^{\dagger}c_{\textbf{\textit{k}}\downarrow}\right\rangle =-\frac{1}{\pi}\sum_{\textbf{\textit{k}}}\int d\omega\frac{{\rm Im}\left[G_{2}\left(\textbf{\textit{k}},\omega\right)\right]}{e^{\omega/T}+1},
\end{equation}
and order parameter equation
\begin{equation}
\frac{\Delta}{U}=-\sum_{\textbf{\textit{k}}}\left\langle c_{-\textbf{\textit{k}}\downarrow}c_{\textbf{\textit{k}}\uparrow}\right\rangle =\frac{1}{\pi}\sum_{\textbf{\textit{k}}}\int d\omega\frac{{\rm Im}\left[\varGamma\left(\textbf{\textit{k}},\omega\right)\right]}{e^{\omega/T}+1},
\end{equation}
with Green's function $G_{1}$, $G_{2}$ and $\varGamma$ in Eq. \ref{eq:GF}
at the temperature $T$. By self-consistently solving the density
and order parameter equations, the value of chemical potential $\mu$
and order parameter $\Delta$ can be numerically calculated.

\begin{figure}
\includegraphics[scale=0.45]{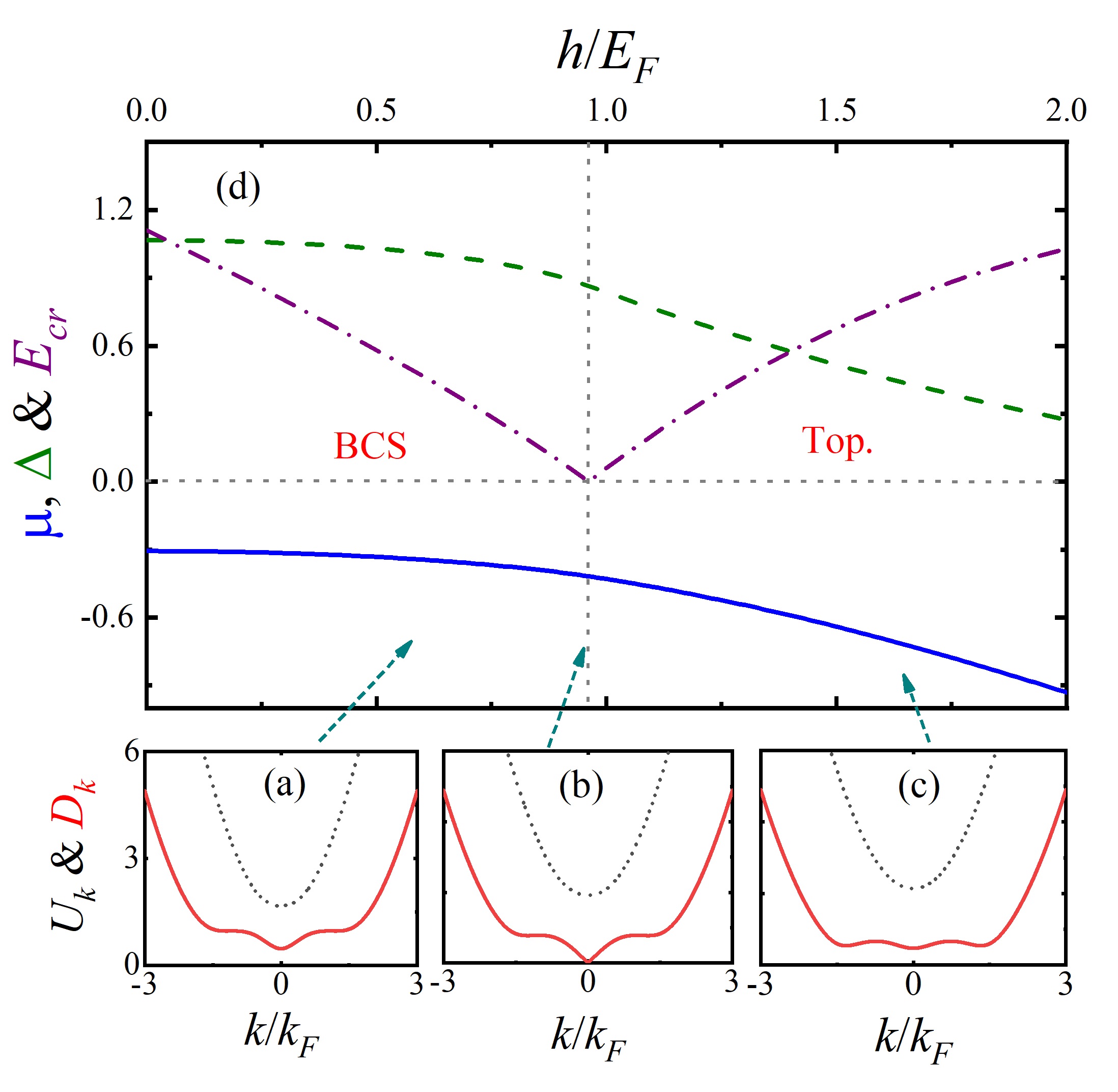}\caption{\label{Fig_transition} Two types of single-particle excitation spectra
$U_{\textbf{\textit{k}}}$ (black short-dotted line) and $D_{\textbf{\textit{k}}}$ (red solid line) at different
Zeeman field (a) $h=0.6E_{F}$, (b) $h=0.96E_{F}$ and (c) $h=1.3E_{F}$.
(d) The distribution of chemical potential (blue solid line), order parameter
(olive dashed line) and $E_{{\rm cr}}=\left|h-\sqrt{\mu^{2}+\Delta^{2}}\right|$ (purple dash-dotted line)
at different Zeeman field $h$ during a continuous phase transition
between BCS superfluid and topological superfluid. A gray vertical dotted
line marks the location of a critical value of Zeeman field $h_{c}=0.96E_{F}$
at $E_{a}=0.5E_{F}$ and $\lambda k_{F}/E_{F}=1.5$. }
\end{figure}

In this paper, we just consider the zero temperature case and take the binding energy $E_{a}=0.5E_{F}$ and SOC
strength $\lambda k_{F}/E_{F}=1.5$. As shown in Fig. \ref{Fig_transition},
the system experiences a phase transition from BCS superfluid to topological
superfluid by increasing the Zeeman field $h$ over a critical value
$h_{c}=0.96E_{F}$. This is a continuous phase transition, which are
displayed by the smooth variation of chemical potential $\mu$ and
order parameter $\Delta$ with Zeeman field $h$. The critical Zeeman
field $h_{c}$ is determined by the zero value of $E_{{\rm cr}}=\left|h-\sqrt{\mu^{2}+\Delta^{2}}\right|$,
and also in which the minimum of $D_{\textbf{\textit{k}}}$ touches zero at momentum
$\textbf{\textit{k}}=0$ ( red line of panel (b) in Fig. \ref{Fig_transition}). During
this continuous phase transition from BCS superfluid to topological one, the  value of the second-order  local minimum in lower-branch spectrum $D_{\textbf{\textit{k}}}$ at a non-zero
momentum $\textbf{\textit{k}}$ will experience a variation from the situation that
much larger than the global minimum at $\textbf{\textit{k}}=0$ in the BCS regime, then to the case
almost the same value of the global one at $\textbf{\textit{k}}=0$ in the topological regime,
while the spectrum structure of $U_{\textbf{\textit{k}}}$ does not change too much.
We have also checked that this continuous phase transition will be
present in a quite large parameter space of SOC strength $\lambda$,
except a weak SOC strength $\lambda k_{F}/E_{F}<0.4$ where the parameter
space of topological superfluid is depressed to almost vanish and make the
phase transition to be a first order one from a trivial superfluid
to normal state.

Next we will discuss the dynamical excitation of this system.

\subsection{Response function and random phase approximation}

At zero temperature, the interacting system comes into a superfluid
state and induces four different densities. Besides the normal spin-up
density $n_{1}=\left\langle \psi_{\uparrow}^{\dagger}\psi_{\uparrow}\right\rangle $
and spin-down density $n_{2}=\left\langle \psi_{\downarrow}^{\dagger}\psi_{\downarrow}\right\rangle $,
the pairing physics of two spins generates the other anomalous density
$n_{3}=\left\langle \psi_{\downarrow}\psi_{\uparrow}\right\rangle $
and its conjugate counterpart $n_{4}=\left\langle \psi_{\uparrow}^{\dagger}\psi_{\downarrow}^{\dagger}\right\rangle $.
The interaction between particles makes these four densities couple
closely with each other. Any fluctuation in each kind of density will
influence other densities and generate a non-negligible density fluctuation
of them. This physics plays a significant role in the dynamical excitation
of the system, which demonstrates the importance and necessity of
the term in Hamiltonian beyond mean-field theory. Random phase approximation
has been verified to be a good way to treat the fluctuation
term of Hamiltonian \citep{AndersonPR1958}. Comparing with experiments,
it can even obtain some quantitatively reliable predictions in three-dimensional
Fermi superfluid \citep{Zou2010q,Zou2018l}. Its prediction also qualitatively
agrees with quantum Monte Carlo data in 2D Fermi system
\citep{Zhao2020d,Vitali2017}. Random phase approximation treats the fluctuation
of Hamiltonian as parts of an effective external potential \citep{Gao2023pra,Liu2004c},
and find the response function $\chi$ of the system is connected
to its mean-field approximation $\chi^{0}$, whose calculation is
relatively easier, by the following equation

\begin{equation}
\chi=\frac{\chi^{0}}{1-\chi^{0}M_{I}U},\label{eq:RPA}
\end{equation}
where
\[
M_{I}=\left[\begin{array}{cccc}
0 & 1 & 0 & 0\\
1 & 0 & 0 & 0\\
0 & 0 & 0 & 1\\
0 & 0 & 1 & 0
\end{array}\right]
\]
 is a constant matrix reflecting the coupling situation of four kinds
of densities.

Next we introduce the expression of the mean-field response function
$\chi^{0}$, which is a $4\times4$ matrix
\begin{equation}
\chi^{0}=\left[\begin{array}{cccc}
\chi_{11}^{0} & \chi_{12}^{0} & \chi_{13}^{0} & \chi_{14}^{0}\\
\chi_{21}^{0} & \chi_{22}^{0} & \chi_{23}^{0} & \chi_{24}^{0}\\
\chi_{31}^{0} & \chi_{32}^{0} & \chi_{33}^{0} & \chi_{34}^{0}\\
\chi_{41}^{0} & \chi_{42}^{0} & \chi_{43}^{0} & \chi_{44}^{0}
\end{array}\right].\label{eq:kai0}
\end{equation}
Here its $ij$ matrix element is defined by $\chi_{ij}^{0}\left(\textbf{\textit{r}}_{1},\textbf{\textit{r}}_{2},\tau,0\right)\equiv-\left\langle \hat{n}_{i}\left(\textbf{\textit{r}}_{1},\tau\right)\hat{n}_{j}\left(\textbf{\textit{r}}_{2},0\right)\right\rangle $, where density operators $\hat{n}_{i}$ and $\hat{n}_{j}$ had been introduced at the beginning of this sub-section.
In the uniform system, all response function elements are only the
function of 2D relative coordinate $\textbf{\textit{r}}=\textbf{\textit{r}}_{1}-\textbf{\textit{r}}_{2}$ and time $\tau$.
So a generalized coordinate $R=\left(\textbf{\textit{r}},\tau\right)$ is used to go
on discussing. Based on Wick's theorem, we should consider all possible
two-operators contraction terms, which are all related to 6 independent
Green's functions of Eq. \ref{eq:GF}. We find that the mean-field
response function $\chi^{0}=A+B$, in which $A$ is only connected to Green's functions $G_{1}$ , $G_{2}$
and $\Gamma$, while $B$ is connected
the SOC Green's functions $S$, $F_{1}$ and $F_{2}$. For example,
in the spatial and time representation, $\chi_{11}^{0}\left(R\right)\equiv-\left\langle \hat{n}_{1}\left(\textbf{\textit{r}}_{1},\tau\right)\hat{n}_{1}\left(\textbf{\textit{r}}_{2},0\right)\right\rangle =A_{11}\left(R\right)+B_{11}\left(R\right)$,
where $A_{11}\left(R\right)=G_{1}\left(-R\right)G_{1}\left(R\right)$
and $B_{11}\left(R\right)=-F_{1}^{*}\left(-R\right)F_{1}\left(R\right)$.
After Fourier transformation to Green's functions, we obtain the expression
of all matrix elements in the momentum-energy representation

\begin{equation}
\chi^{0}\left(\textbf{\textit{q}},\omega\right)=A\left(\textbf{\textit{q}},\omega\right)+B\left(\textbf{\textit{q}},\omega\right)\label{eq:xab}
\end{equation}
where
\[
A=\left[\begin{array}{cccc}
A_{11}, & A_{12}, & A_{13}, & A_{14}\\
A_{12}, & A_{22}, & A_{23}, & A_{24}\\
A_{14}, & A_{24}, & -A_{12}, & A_{34}\\
A_{13}, & A_{23}, & A_{43}, & -A_{12}
\end{array}\right]
\]
has 9 independent matrix elements, and

\[
B=\left[\begin{array}{cccc}
B_{11}, & B_{12}, & B_{13}, & B_{14}\\
B_{21}, & B_{22}, & B_{23}, & B_{24}\\
B_{31}, & B_{32}, & B_{33}, & B_{34}\\
B_{41}, & B_{42}, & B_{43}, & B_{44}
\end{array}\right].
\]
 All expressions of these matrix elements are listed in the final
appendix. The numerical calculation of above all matrix elements required
a two-dimensional integral, which makes the numerical calculations here
much heavier than the one-dimensional SOC system \citep{Gao2023pra}.

\subsection{Dynamic structure factor}

With Eqs. \ref{eq:RPA} and \ref{eq:xab}, we can obtain expression
of both the total density response function $\chi_{n}\equiv\chi_{11}+\chi_{22}+\chi_{12}+\chi_{21}$
and spin density response function $\chi_{s}\equiv\chi_{11}+\chi_{22}-\chi_{12}-\chi_{21}$.
$\chi_{n}$ reflects the density response of the system, while
$\chi_{s}$ shows the spin density response in two-spin components.
Based on the fluctuation and dissipation theorem, their imaginal parts
are connected to density and spin dynamic structure factor by

\begin{equation}
S_{n/s}=-\frac{1}{\pi}\frac{1}{1-e^{-\omega/T}}{\rm Im}\left[\chi_{n/s}\right].
\end{equation}

\section{Results}

\begin{figure}
\includegraphics[scale=0.35]{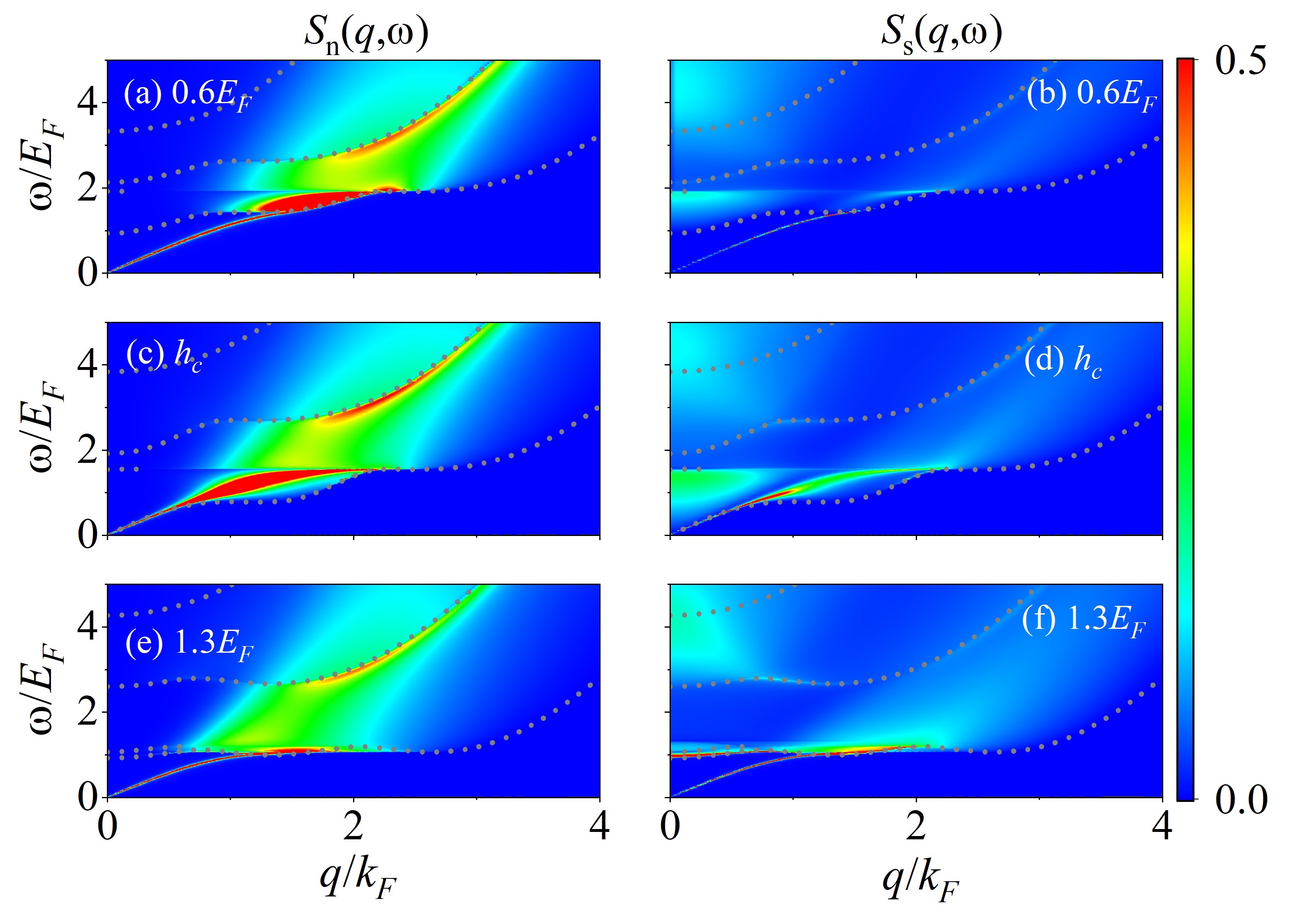}\caption{\label{Fig_dsf} The density (left column) and spin (right column)
dynamic structure factor at three typical different Zeeman fields
$h=0.6E_{F},h_{c},1.3E_{F}$. The dotted lines are the same lines of Fig. \ref{Fig_spectrum}, which reflects all kinds of threshold energy to break a Cooper pair.}
\end{figure}

In the following discussions, we focus on an interaction binding
energy $E_{a}=0.5E_{F}$ and a typical SOC strength $\lambda k_{F}/E_{F}=1.5$
at zero temperature. These parameters are the same as one in Fig.
\ref{Fig_transition}. As introduced before, the isotropy of the Rashba SOC effect induces that the Hamiltonian Eq. \ref{eq_Ham} is also isotropic, which means the direction of the transferred momentum $\textbf{\textit{q}}$ can make no difference to the dynamical excitation of the system, so we just set $\textbf{\textit{q}}$ along the positive direction of $X$-axis. And the dynamical excitation of the system is also rotation-invariant in the 2D $XY$-plane.

We numerically calculate the density (left
column) and spin (right column) dynamic structure factors, as shown
in Fig. \ref{Fig_dsf}, in the phase transition from BCS superfluid
(higher two panels), cross the critical regime (middle two panels),
and then to topological superfluid (lower two panels). Generally
we investigate a full dynamical excitation in different transferred
momenta $\textbf{\textit{q}}$, including the low energy (or momentum) collective excitation
and the high energy (or momentum) single-particle excitation. The
white dotted lines mark the location of three types of the minimum energy to break a Cooper pair, which will be introduced later.

\subsection{Collective and single-particle excitation}

\begin{figure}
\includegraphics[scale=0.3]{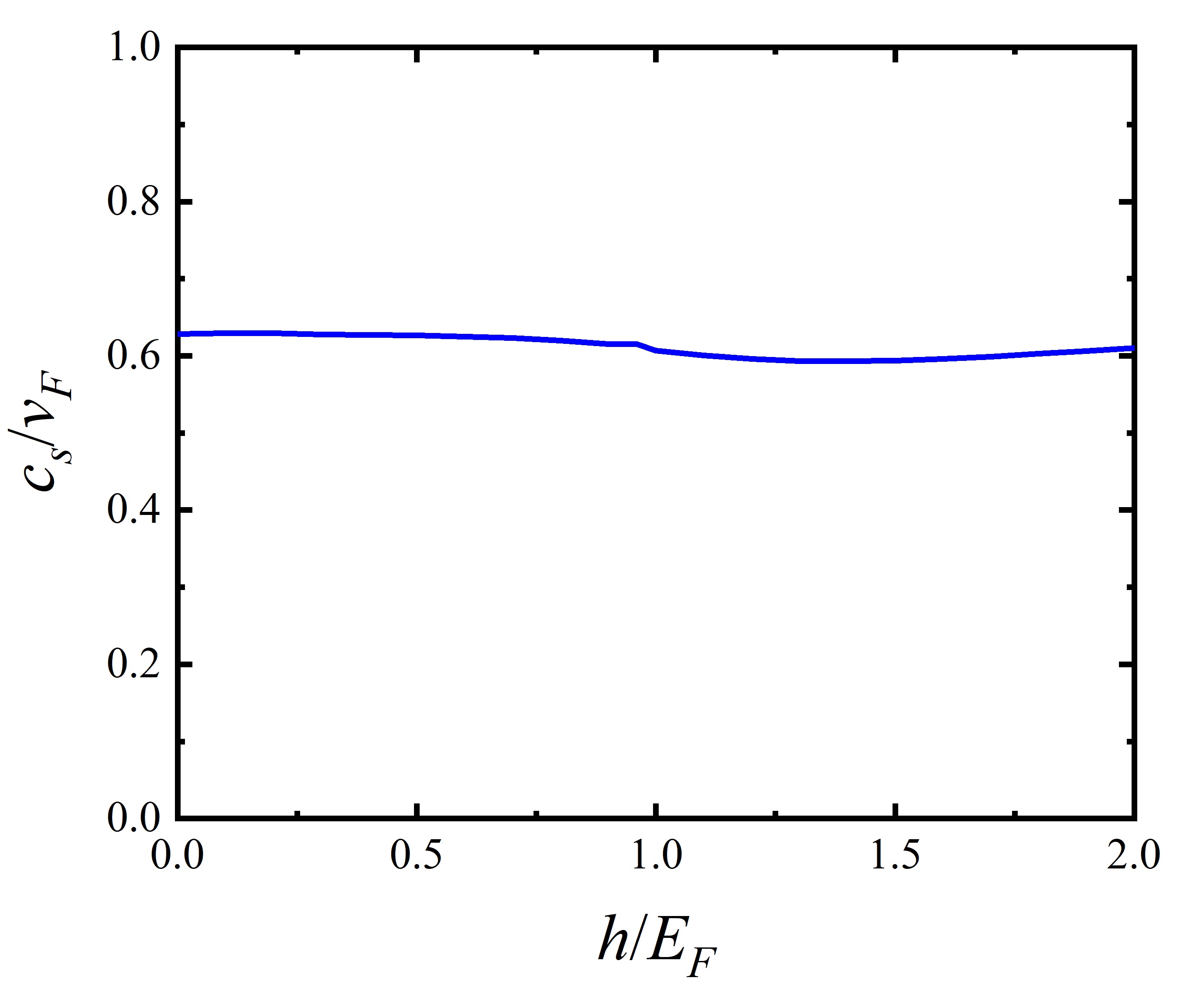}\caption{\label{Fig_sound} The sound velocity $c_s$ at different Zeeman field
$h$.}
\end{figure}

At a low transferred energy $\omega$, it is easy to investigate the
collective excitation \citep{He2013aop}. By continuously increasing transferred momentum $\textbf{\textit{q}}$ from zero, we initially see a gapless phonon excitation in both
the density and spin dynamic structure factor. As shown in Fig. \ref{Fig_sound},
the velocity of collective phonon excitation almost does not change
during this continuous phase transition, which is different from the first order one
in 1D Raman SOC Fermi superfluid \citep{Gao2023pra}. When the transferred
momentum $\textbf{\textit{q}}$ is large enough, this phonon excitation gradually merges
into the continuous single-particle excitation. Specially in the critical
regime $h=h_{c}$, the minimum of lower single-particle spectrum $D_{\textbf{\textit{k}}}$
touches zero (panel (b) of Fig. \ref{Fig_transition}), which induces
that a gapless pair-breaking excitation happens at the same location of
collective phonon excitation and give a finite expansion width to
phonon excitation.

\begin{figure}
\includegraphics[scale=0.34]{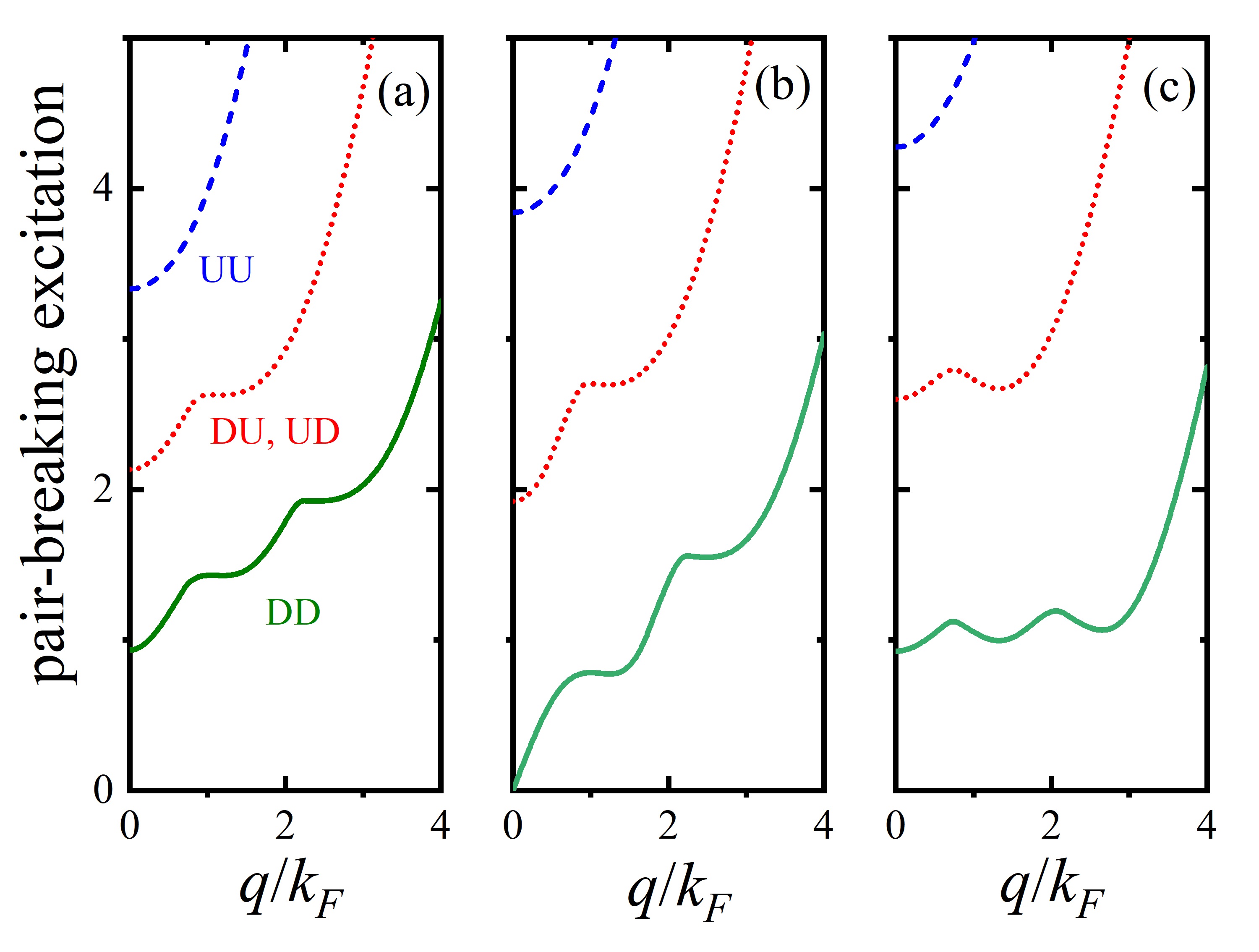}\caption{\label{Fig_spectrum} All kinds of threshold energy of pair-breaking
excitation in different momenta at (a) $h=0.6E_F$, (b) $h=0.96E_F$ and (c) $h=1.3E_F$. Olive solid line: $D_{\textbf{\textit{k}}}\leftrightarrow D_{\textbf{\textit{k}}+\textbf{\textit{q}}}.$
Red dotted line: $D_{\textbf{\textit{k}}}\leftrightarrow U_{\textbf{\textit{k}}+\textbf{\textit{q}}}$ and $U_{\textbf{\textit{k}}}\leftrightarrow D_{\textbf{\textit{k}}+\textbf{\textit{q}}}$.
Blue dashed line: $U_{\textbf{\textit{k}}}\leftrightarrow U_{\textbf{\textit{k}}+\textbf{\textit{q}}}$.}
\end{figure}

When the transferred energy $\omega$ is large enough, the excitation
of the system is dominated by the sing-particle excitation. A pair-breaking
of Cooper pairs will occur and make pairs be broken into free Fermi
atoms. Indeed a great regime of the dynamical excitation in Fig. \ref{Fig_dsf}
is dominated by this pair-breaking effect. In the density dynamic
structure factor $S_{n}$, this effect usually is very obvious in
a relatively large transferred momentum $\textbf{\textit{q}}>k_{F}$, where the collective
excitation is strongly depressed. Different from the conventional
Fermi superfluid, this single-particle excitation takes up a large
regime in the spin dynamic structure factor $S_{s}$, even for a small
or zero transferred momentum $\textbf{\textit{q}}$. To understand this, it is necessary to study the threshold energy to break
a Cooper pair. This pair-breaking excitation is related to two branches
of quasi-particle spectra $U_{\textbf{\textit{k}}}$ and $D_{\textbf{\textit{k}}}$. The two atoms
forming a Cooper-pair can come from the same or opposite branch of
single particle spectrum. This two-branch spectrum structure generates
four kinds of mechanism to break a Cooper pair, namely the $DD$,
$DU$, $UD$, and $UU$ type. The minimum energy at a certain momentum
$\textbf{\textit{q}}$ to break a pair is ${\rm min}[D_{\textbf{\textit{k}}}+D_{\textbf{\textit{k}}+\textbf{\textit{q}}}]$, ${\rm min}[D_{\textbf{\textit{k}}}+U_{\textbf{\textit{k}}+\textbf{\textit{q}}}]$,
${\rm min}[U_{\textbf{\textit{k}}}+D_{\textbf{\textit{k}}+\textbf{\textit{q}}}]$ or ${\rm min}[U_{\textbf{\textit{k}}}+U_{\textbf{\textit{k}}+\textbf{\textit{q}}}]$. Here the
$DU$ and $UD$ excitations are overlapped with each other, and finally
display three kinds of pair-breaking excitation regime. The minimum
energy in these pair-breaking excitation are shown by three panels
of Fig. \ref{Fig_spectrum}, which are also displayed by the dotted
lines in Fig. \ref{Fig_dsf}. The lowest olive line denotes the $DD$ type minimum energy (${\rm min}[D_{\textbf{\textit{k}}}+D_{\textbf{\textit{k}}+\textbf{\textit{q}}}]$) to break a Cooper pair at a certain $\textbf{\textit{q}}$, and atoms forming a Cooper pair are both from
the down-branch quasi-particle spectrum $D_\textbf{\textit{k}}$.
Generally this value is always larger than zero in both BCS superfluid
(panel (a)) and topological superfluid (panel (c)). However it will touch
zero at the critical Zeeman field $h=0.96E_F$ (panel (b)), which is an important
signal of this continuous phase transition. The red line denotes the minimum energy of cross-spectrum
excitation ($DU$ and $UD$ type). The two atoms in a pair come from
different branches of spectrum. This excitation starts from the ${\rm min}[D_{\textbf{\textit{k}}}+U_{\textbf{\textit{k}}+\textbf{\textit{q}}}]$,
and it requires an energy higher than the $DD$ one. This cross-spectrum
excitation also reflects the coupling effect between spin- and orbital
motion, and it is much easier to be observed in the spin dynamic structure
factor than that in density one. The blue dashed line is the minimum energy in
$UU$ excitation. Its energy is the largest among three kinds of pair-breaking excitation, however the excitation signal of $UU$ excitation is the weakest.

Besides all global minima discussed above, there are also some possible
local minima in these pair-breaking excitations, which generate some
edges in the dynamic structure factor. For example, some horizon edges
emerge when $\omega$ is a little lower than $2E_{F}$ in the dynamic structure factor of Fig. \ref{Fig_dsf}, which is from the local minimum of the $DD$ type pair-breaking excitation.

To better understand the dynamical excitation in these colorful panels,
we also discuss the dynamic structure factor at a fixed transferred
momentum $\textbf{\textit{q}}$.

\subsection{Dynamic excitation at a constant momentum $\textbf{\textit{q}}$}

\begin{figure}
\includegraphics[scale=0.3]{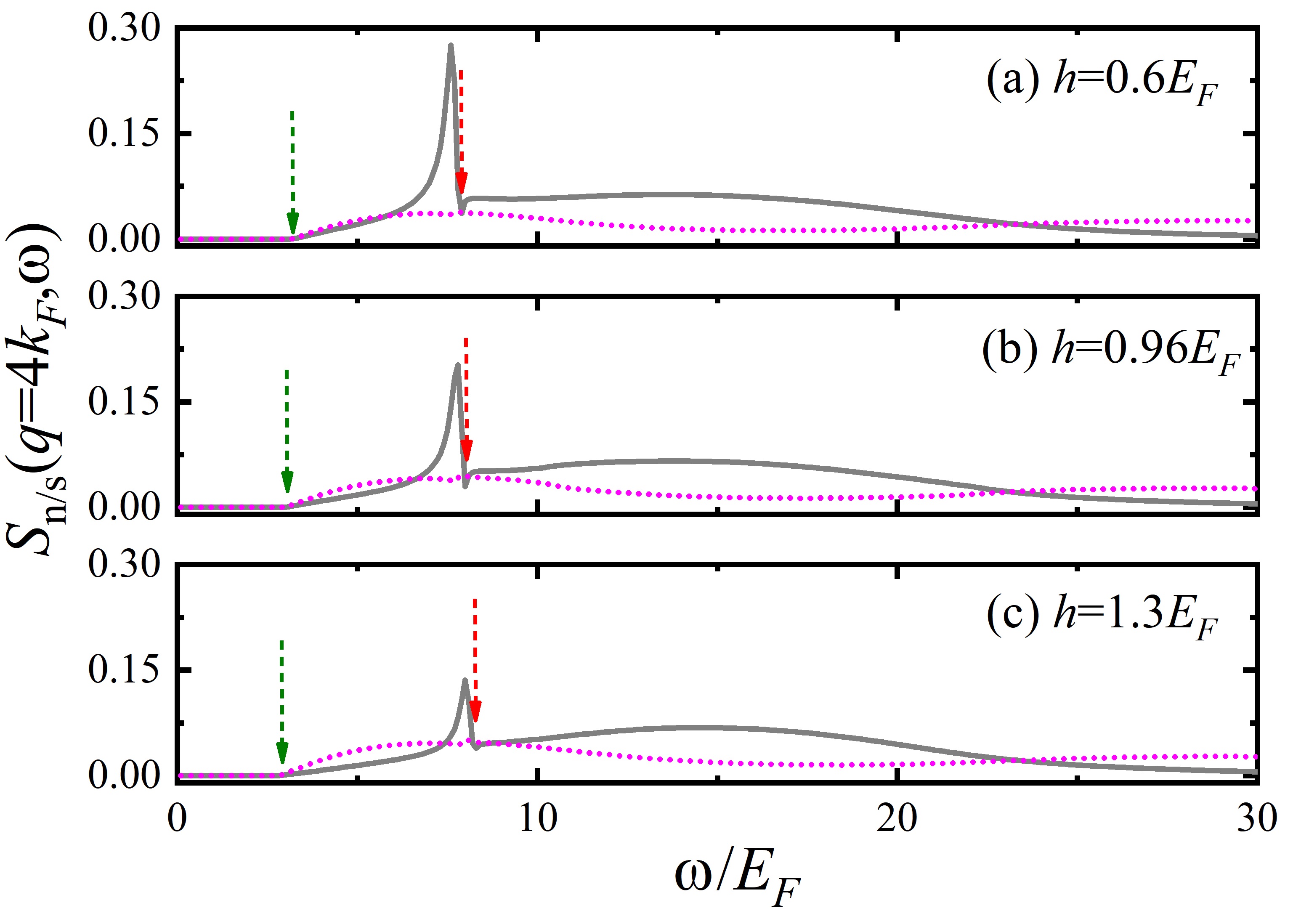}\caption{\label{Fig_q4} The density (gray) and spin (magenta) dynamical structure
factor at transferred momentum $\textbf{\textit{q}}=4k_{F}$. (a) $h=0.6E_{F}$ at BCS
superfluid, (b) $h=0.96E_{F}$ at transition point, (c) $h=1.3E_{F}$
at topological superfluid.}
\end{figure}

For a large transferred momentum $\textbf{\textit{q}}\gg k_{F}$, the dynamic structure
factor is dominated by the single-particle excitation. As shown in
Fig. \ref{Fig_q4}, we investigate the density and spin dynamic structure
factors at $\textbf{\textit{q}}=4k_{F}$ between BCS and topological superfluid. In all
three panels, we always find a high excitation signal in density dynamic
structure factor (gray solid lines) around $\omega=8E_{F}$. In fact
it is the molecular Cooper-pair excitation, whose dispersion relation
can be easily explained by $\varepsilon_{k}=\textbf{\textit{q}}^{2}/2M$ and $M=2m$
is the mass a two-atom molecule. Also the single-atom excitation arrives
its maximum around $\textbf{\textit{q}}^{2}/2m\thickapprox16E_{F}$ here. The olive
and red arrows respectively mark the threshold energy to break a
Cooper pair in $DD$ and $DU$ (or $UD$) type excitation. Different from
the 3D crossover Fermi superfluid \citep{Combescot2006m}, the Rashba-SOC
effect makes $DD$ pair-breaking excitation happen earlier than molecular
excitation, no matter the value of Zeeman field $h$.

\begin{figure}
\includegraphics[scale=0.3]{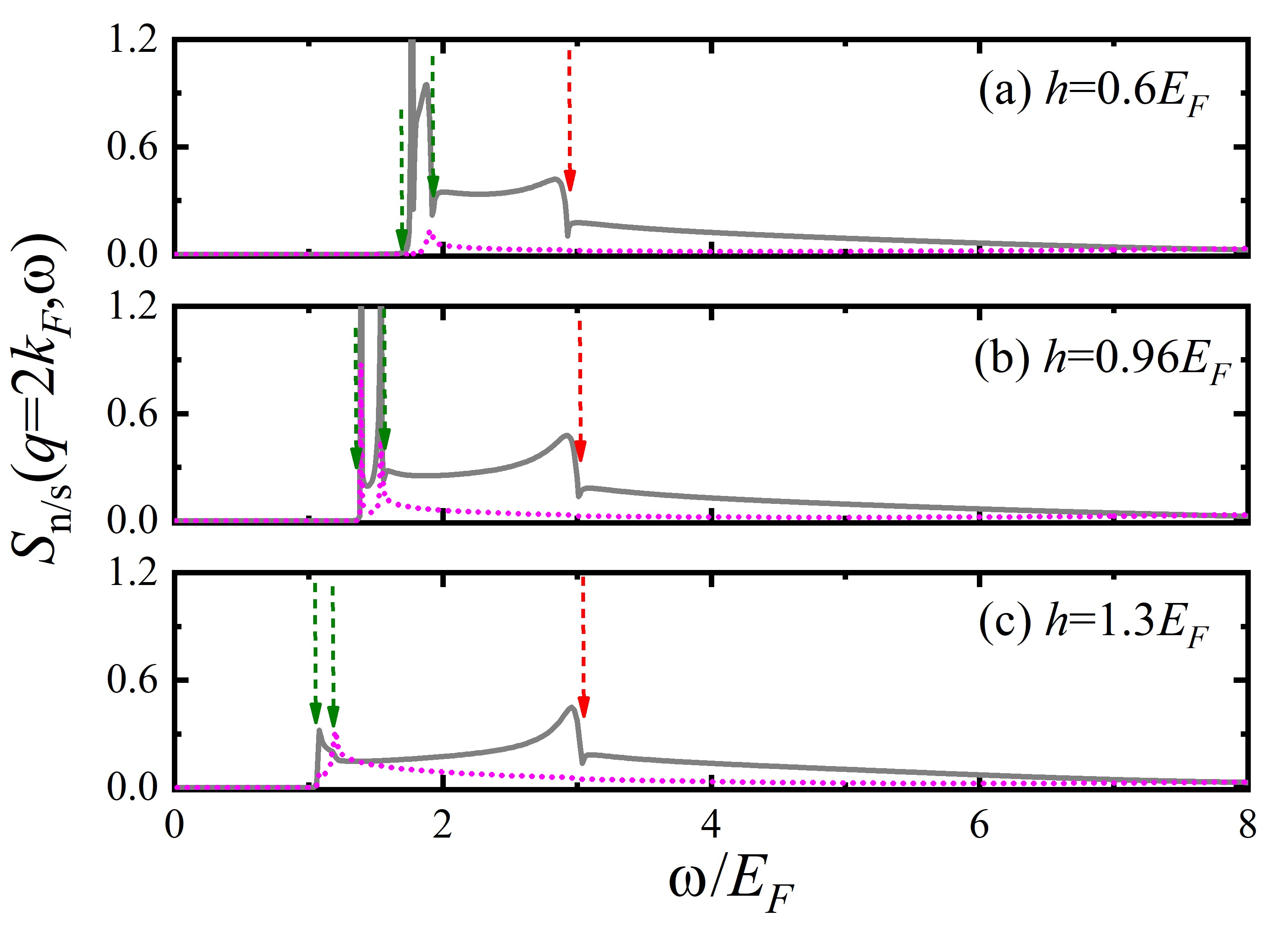}\caption{\label{Fig_q2} The density (gray) and spin (magenta) dynamical structure
factor at transferred momentum $\textbf{\textit{q}}=2k_{F}$. The arrangement of parameters in these three panels is the same as that in Fig. \ref{Fig_q4}.}
\end{figure}

When taking transferred momentum $\textbf{\textit{q}}=2k_{F}$, the collective phonon
has already merged into the regime of single-particle excitation.
The dynamic structure factor is dominated by strong signals of pair-breaking
excitation. As shown in Fig. \ref{Fig_q2}, both curves of $S_{n}$
and $S_{s}$ have many twists which means they exhibit rich oscillations, and the two-olive arrows respectively
mark the global (left) and local (right) minimum energy to break a
Cooper pair according to $DD$ type excitation, and one red-dashed
arrow marks the minimum energy of $DU$ (or $UD$) type excitation.
In all three panels, small peaks (left side of red arrow) in density
dynamic structure factor display the two-atom molecule excitation
around $\omega=2.9E_{F}$, the obvious deviation from dispersion line
$\varepsilon_{k}=\textbf{\textit{q}}^{2}/2M=4E_{F}$ is due to its coupling effect to
pair-breaking excitation in this relative weak transferred momentum
($\textbf{\textit{q}}=2k_{F}$).

\begin{figure}
\includegraphics[scale=0.3]{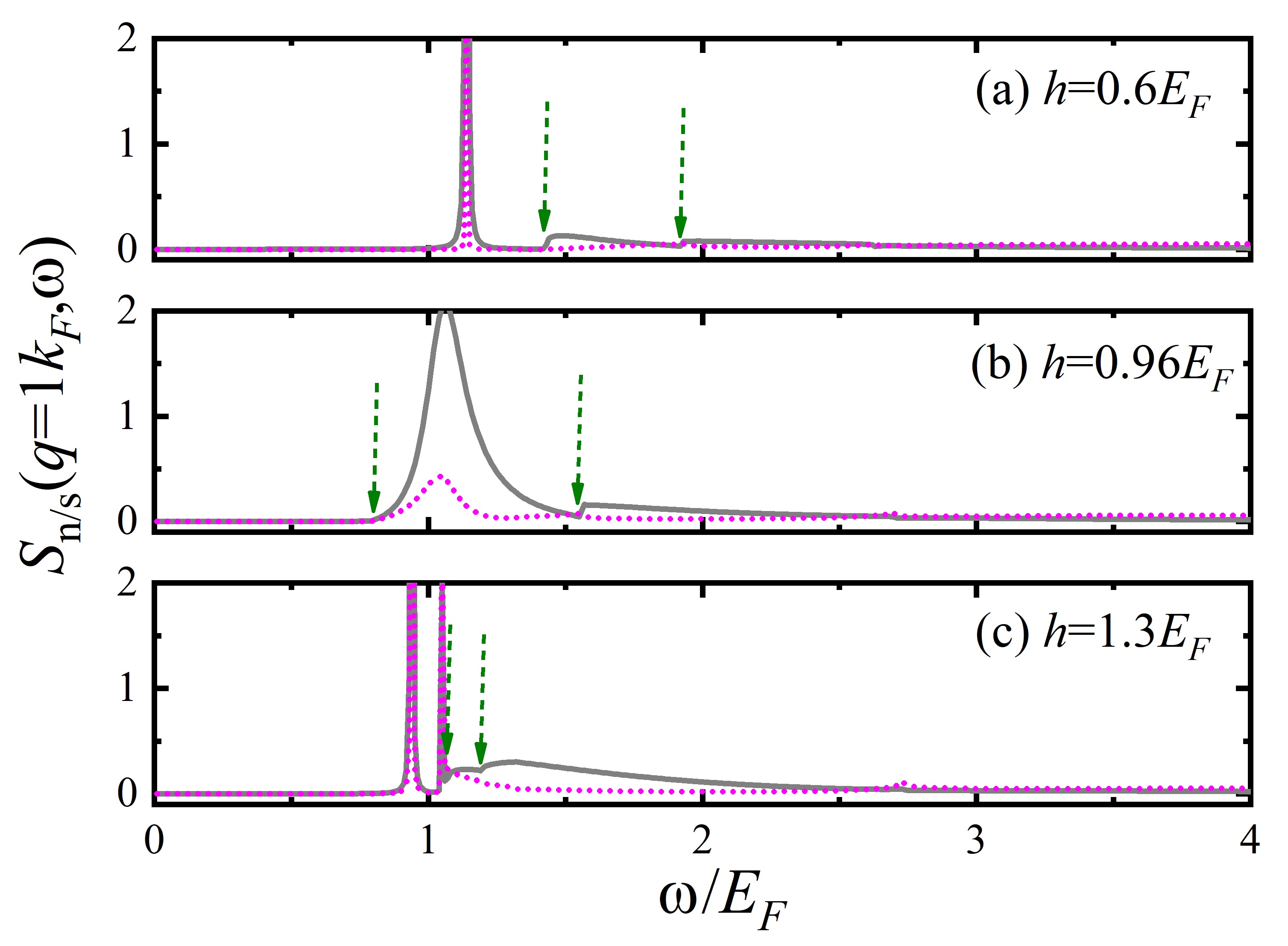}\caption{\label{Fig_q1} The density (gray) and spin (magenta) dynamical structure
factor at transferred momentum $\textbf{\textit{q}}=1k_{F}$. The arrangement of parameters in these three panels is the same as that in Fig. \ref{Fig_q4}.}
\end{figure}

\begin{figure}
\includegraphics[scale=0.3]{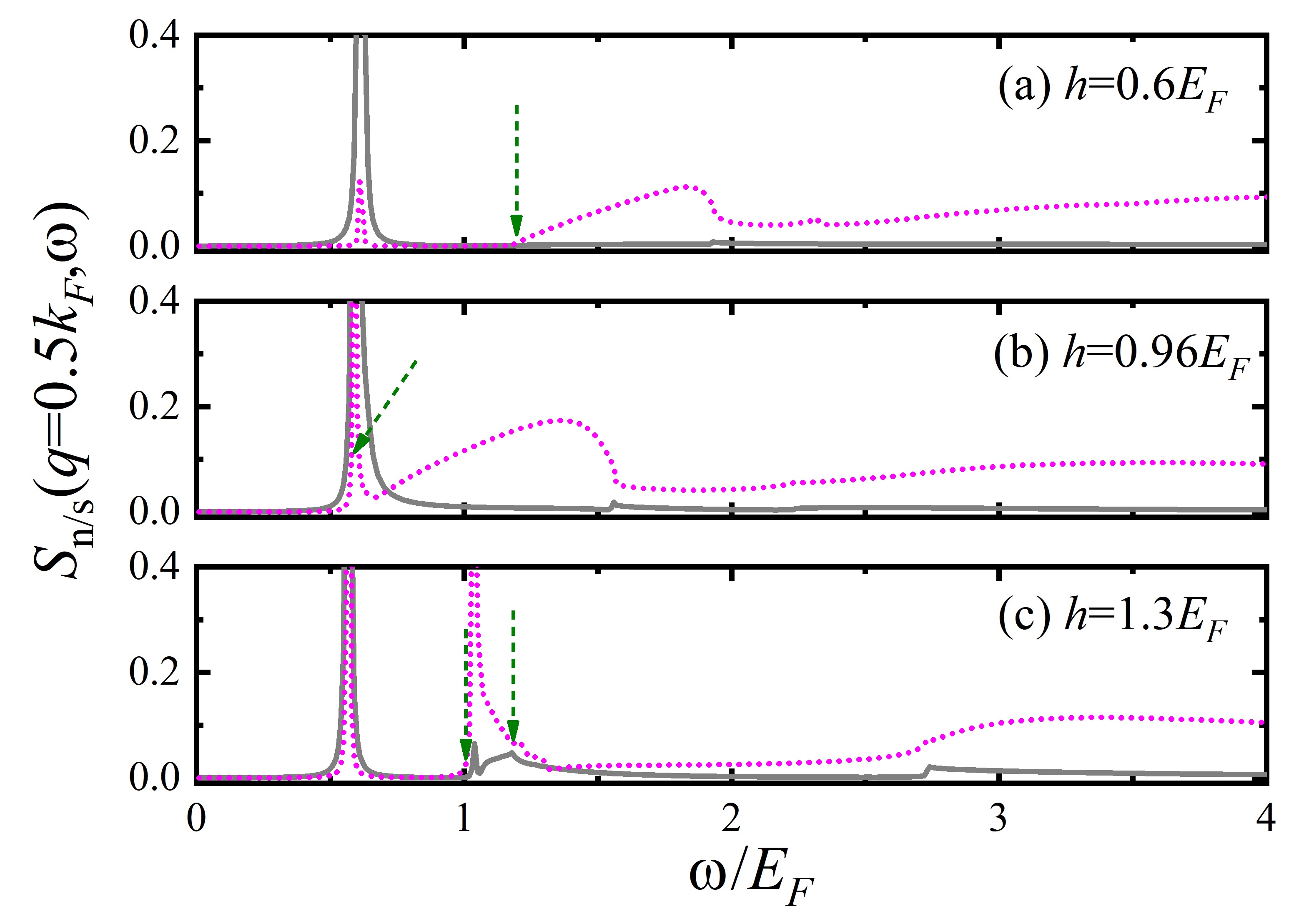}\caption{\label{Fig_q05} The density (gray) and spin (magenta) dynamical structure
factor at transferred momentum $\textbf{\textit{q}}=0.5k_{F}$. The arrangement of parameters in these three panels is the same as that in Fig. \ref{Fig_q4}.}
\end{figure}

For a transferred momentum $\textbf{\textit{q}}$ at the order of Fermi wave vector
$k_{F}$ (or smaller than $k_{F}$), the collective phonon excitation
can be separated from the pair-breaking excitation, and happens at
a smaller energy than pair-breaking effect. At the panel (a) and
(c) of Fig. \ref{Fig_q1} and \ref{Fig_q05}, strong sharp peaks
are shown in the most left side of the dynamic structure factor, whose
excitation energy are smaller than the global minimum energy of $DD$
type pair-breaking excitation. And the separation energy between phonon
excitation and $DD$ type threshold excitation is relatively small
in the topological superfluid ($h=1.3E_{F}$) since it has a weaker
order parameter $\Delta$ than that in the BCS superfluid ($h=0.6E_{F}$).
However in the point of phase transition $h=0.96E_{F}$ (shown by
panel (b) of Figs. \ref{Fig_q1} and \ref{Fig_q05}), the phonon excitation is just overlapped
with the beginning of the gapless $DD$ type pair-breaking excitation, and give a finite width to the phonon peak at $\textbf{\textit{q}}=1k_{F}$ (panel (b) of Fig. \ref{Fig_q1}).
At $\textbf{\textit{q}}=0.5k_{F}$, although these two excitations are still mixed
with each other at $h=0.96E_{F}$, this small transferred momentum
$\textbf{\textit{q}}$ generates a relative weaker strength of $DD$ excitation.
The phonon excitation is still very sharp in the density dynamic structure
factor, and the spin dynamic structure factor can do help to track the
signal of $DD$ excitation and display a bump structure closely following
the phonon peak.

\section{Conclusions and outlook}

In summary, we numerically calculate the density and spin dynamic
structure factor of 2D Rashba SOC Fermi superfluid with random phase
approximation during a continuous phase transition between BCS and
topological superfluids. The dynamic structure factor presents rich
excitation signals, including collective phonon excitation, molecular or atomic excitations, and pair-breaking
excitations. The gapless collective
phonon excitation requires the smallest excitation energy in both BCS
and topological superfluid. In the critical point of phase transition, the phonon excitation is overlapped with a gapless $DD$ type pair-breaking excitation, and also is imparted a finite expansion width to the phonon peak, which is a delta-like peak when far from critical point of phase transition. For a larger transferred momentum $\textbf{\textit{q}}$, the strength of
phonon excitation gradually decreases and merges into the pair-breaking
excitation regime, and the excitation signals in both density and
spin dynamic structure factor are dominated by single-particle excitation, including three kinds of pair-breaking excitation, two-atom molecular and single atomic excitations.
The two-atom molecular (single atomic) excitations can be well explained
by an ideal single molecule (atom) dispersion relation at a very large
transferred momentum $\textbf{\textit{q}} \gg k_{F}$. Our research about dynamic structure factor can do help to understand the dynamical excitation information in both BCS and topological matter state, distinguish different matter state during phase transition and judge the location of phase transition.

In the near future, it will be interesting
to bring some non-uniform structures, like edge, impurity,
or soliton (vortex), to this system, by which it is expected to investigate
some excitations related to the generation of Majorana fermions \citep{Majorana1937nc,Liu2013,Xu2014,Liu2015},
which is absent in the current work. Experimentally the edge can be brought by a hard wall or a harmonic trap, and soliton can be generated by a phase-imprinting technique \citep{KongPRA2021}. So it will be worth to carry out calculation to these non-uniform systems.

\section{Acknowledgements}

This research was supported by the National Natural Science Foundation
of China, Grants No. 11804177 (P.Z.), No. 11547034 (H.Z.), No. 11974384
(S.-G.P.); NKRDP under Grant No. 2016YFA0301503 (S.-G.P.).

\section{Appendix}
The exact diagonalization of mean-field Hamiltonian $H_{{\rm mf}}$ is carried out by motion equation of Green's function $\omega\left\langle \left\langle c_{1}|c_{2}\right\rangle \right\rangle =\left\langle \left[c_{1},c_{2}\right]_{+}\right\rangle +\left\langle \left\langle \left[c_{1},H_{{\rm mf}}\right]|c_{2}\right\rangle \right\rangle $,
where $c_{1}$ and $c_{2}$ are any possible fermionic operators of the system, and the double-bracket notation $\left\langle \left\langle c_{1}|c_{2}\right\rangle \right\rangle$ is the corresponding momentum-energy Fourier transformation of space-time Green's function $G(r_1,\tau,r_2,0)=-\left<T\psi_{1}(r_1,\tau)\psi_{2}(r_2,0)\right>$. Finally we find that the system has six independent Green's functions. In this appendix, we will list expressions of six independent Green's
functions and mean-field response function $\chi^{0}=A+B$.
The six independent Green's functions are $G_{1}\left(\textbf{\textit{k}},\omega\right)=\sum_{l}\left[G_{1}\right]_{\textbf{\textit{k}}}^{l}/\left(\omega-E_{\textbf{\textit{k}}}^{l}\right)$
with
\begin{flushleft}
\[
\begin{array}{cl}
\left[G_{1}\right]_{\textbf{\textit{k}}}^{(1)} & =\frac{U_{\textbf{\textit{k}}}^{2}-\left(h^{2}+\textbf{\textit{k}}^{2}\lambda^{2}+E_{\textbf{\textit{k}}}^{2}+2h\xi_{\textbf{\textit{k}}}\right)}{2\left(U_{\textbf{\textit{k}}}^{2}-D_{\textbf{\textit{k}}}^{2}\right)}\\
 & +\frac{\left(\xi_{\textbf{\textit{k}}}-h\right)U_{\textbf{\textit{k}}}^{2}-\left(\xi_{\textbf{\textit{k}}}+h\right)\left(E_{\textbf{\textit{k}}}^{2}-h^{2}-\textbf{\textit{k}}^{2}\lambda^{2}\right)}{2U_{\textbf{\textit{k}}}\left(U_{\textbf{\textit{k}}}^{2}-D_{\textbf{\textit{k}}}^{2}\right)},
\end{array}
\]
\par\end{flushleft}

\begin{flushleft}
\[
\begin{array}{cl}
\left[G_{1}\right]_{\textbf{\textit{k}}}^{(-1)} & =\frac{U_{\textbf{\textit{k}}}^{2}-\left(h^{2}+\textbf{\textit{k}}^{2}\lambda^{2}+E_{\textbf{\textit{k}}}^{2}+2h\xi_{\textbf{\textit{k}}}\right)}{2\left(U_{\textbf{\textit{k}}}^{2}-D_{\textbf{\textit{k}}}^{2}\right)}\\
 & -\frac{\left(\xi_{\textbf{\textit{k}}}-h\right)U_{\textbf{\textit{k}}}^{2}-\left(\xi_{\textbf{\textit{k}}}+h\right)\left(E_{\textbf{\textit{k}}}^{2}-h^{2}-\textbf{\textit{k}}^{2}\lambda^{2}\right)}{2U_{\textbf{\textit{k}}}\left(U_{\textbf{\textit{k}}}^{2}-D_{\textbf{\textit{k}}}^{2}\right)},
\end{array}
\]
\par\end{flushleft}

\begin{flushleft}
\[
\begin{array}{cl}
\left[G_{1}\right]_{\textbf{\textit{k}}}^{(2)} & =-\frac{D_{\textbf{\textit{k}}}^{2}-\left(h^{2}+\textbf{\textit{k}}^{2}\lambda^{2}+E_{\textbf{\textit{k}}}^{2}+2h\xi_{\textbf{\textit{k}}}\right)}{2\left(U_{\textbf{\textit{k}}}^{2}-D_{\textbf{\textit{k}}}^{2}\right)}\\
 & -\frac{\left(\xi_{\textbf{\textit{k}}}-h\right)D_{\textbf{\textit{k}}}^{2}-\left(\xi_{\textbf{\textit{k}}}+h\right)\left(E_{\textbf{\textit{k}}}^{2}-h^{2}-\textbf{\textit{k}}^{2}\lambda^{2}\right)}{2D_{\textbf{\textit{k}}}\left(U_{\textbf{\textit{k}}}^{2}-D_{\textbf{\textit{k}}}^{2}\right)},
\end{array}
\]
\par\end{flushleft}

\begin{flushleft}
\[
\begin{array}{cl}
\left[G_{1}\right]_{\textbf{\textit{k}}}^{(-2)} & =-\frac{D_{\textbf{\textit{k}}}^{2}-\left(h^{2}+\textbf{\textit{k}}^{2}\lambda^{2}+E_{\textbf{\textit{k}}}^{2}+2h\xi_{\textbf{\textit{k}}}\right)}{2\left(U_{\textbf{\textit{k}}}^{2}-D_{\textbf{\textit{k}}}^{2}\right)}\\
 & +\frac{\left(\xi_{\textbf{\textit{k}}}-h\right)D_{\textbf{\textit{k}}}^{2}-\left(\xi_{\textbf{\textit{k}}}+h\right)\left(E_{\textbf{\textit{k}}}^{2}-h^{2}-\textbf{\textit{k}}^{2}\lambda^{2}\right)}{2D_{\textbf{\textit{k}}}\left(U_{\textbf{\textit{k}}}^{2}-D_{k}^{2}\right)}.
\end{array}
\]
\par\end{flushleft}

$G_{2}\left(\textbf{\textit{k}},\omega\right)=\sum_{l}\left[G_{2}\right]_{\textbf{\textit{k}}}^{l}/\left(\omega-E_{\textbf{\textit{k}}}^{l}\right)$
with
\begin{flushleft}
\[
\begin{array}{cl}
\left[G_{2}\right]_{\textbf{\textit{k}}}^{(1)} & =\frac{U_{\textbf{\textit{k}}}^{2}-\left(h^{2}+\textbf{\textit{k}}^{2}\lambda^{2}+E_{\textbf{\textit{k}}}^{2}-2h\xi_{\textbf{\textit{k}}}\right)}{2\left(U_{\textbf{\textit{k}}}^{2}-D_{\textbf{\textit{k}}}^{2}\right)}\\
 & +\frac{\left(\xi_{\textbf{\textit{k}}}+h\right)U_{\textbf{\textit{k}}}^{2}-\left(\xi_{\textbf{\textit{k}}}-h\right)\left(E_{\textbf{\textit{k}}}^{2}-h^{2}-\textbf{\textit{k}}^{2}\lambda^{2}\right)}{2U_{\textbf{\textit{k}}}\left(U_{\textbf{\textit{k}}}^{2}-D_{\textbf{\textit{k}}}^{2}\right)},
\end{array}
\]
\par\end{flushleft}

\begin{flushleft}
\[
\begin{array}{cl}
\left[G_{2}\right]_{\textbf{\textit{k}}}^{(-1)} & =\frac{U_{\textbf{\textit{k}}}^{2}-\left(h^{2}+\textbf{\textit{k}}^{2}\lambda^{2}+E_{\textbf{\textit{k}}}^{2}-2h\xi_{\textbf{\textit{k}}}\right)}{2\left(U_{\textbf{\textit{k}}}^{2}-D_{\textbf{\textit{k}}}^{2}\right)}\\
 & -\frac{\left(\xi_{\textbf{\textit{k}}}+h\right)U_{\textbf{\textit{k}}}^{2}-\left(\xi_{\textbf{\textit{k}}}-h\right)\left(E_{\textbf{\textit{k}}}^{2}-h^{2}-\textbf{\textit{k}}^{2}\lambda^{2}\right)}{2U_{\textbf{\textit{k}}}\left(U_{\textbf{\textit{k}}}^{2}-D_{\textbf{\textit{k}}}^{2}\right)},
\end{array}
\]
\par\end{flushleft}

\begin{flushleft}
\[
\begin{array}{cl}
\left[G_{2}\right]_{\textbf{\textit{k}}}^{(2)} & =-\frac{D_{\textbf{\textit{k}}}^{2}-\left(h^{2}+\textbf{\textit{k}}^{2}\lambda^{2}+E_{\textbf{\textit{k}}}^{2}-2h\xi_{\textbf{\textit{k}}}\right)}{2\left(U_{\textbf{\textit{k}}}^{2}-D_{\textbf{\textit{k}}}^{2}\right)}\\
 & -\frac{\left(\xi_{\textbf{\textit{k}}}+h\right)D_{\textbf{\textit{k}}}^{2}-\left(\xi_{\textbf{\textit{k}}}-h\right)\left(E_{\textbf{\textit{k}}}^{2}-h^{2}-\textbf{\textit{k}}^{2}\lambda^{2}\right)}{2D_{\textbf{\textit{k}}}\left(U_{\textbf{\textit{k}}}^{2}-D_{\textbf{\textit{k}}}^{2}\right)},
\end{array}
\]
\par\end{flushleft}

\begin{flushleft}
\[
\begin{array}{cl}
\left[G_{2}\right]_{\textbf{\textit{k}}}^{(-2)} & =-\frac{D_{\textbf{\textit{k}}}^{2}-\left(h^{2}+\textbf{\textit{k}}^{2}\lambda^{2}+E_{\textbf{\textit{k}}}^{2}-2h\xi_{\textbf{\textit{k}}}\right)}{2\left(U_{\textbf{\textit{k}}}^{2}-D_{\textbf{\textit{k}}}^{2}\right)}\\
 & +\frac{\left(\xi_{\textbf{\textit{k}}}+h\right)D_{\textbf{\textit{k}}}^{2}-\left(\xi_{\textbf{\textit{k}}}-h\right)\left(E_{\textbf{\textit{k}}}^{2}-h^{2}-\textbf{\textit{k}}^{2}\lambda^{2}\right)}{2D_{\textbf{\textit{k}}}\left(U_{\textbf{\textit{k}}}^{2}-D_{\textbf{\textit{k}}}^{2}\right)},
\end{array}
\]
\par\end{flushleft}

$\varGamma\left(\textbf{\textit{k}},\omega\right)=\sum_{l}\left[\varGamma\right]_{\textbf{\textit{k}}}^{l}/\left(\omega-E_{\textbf{\textit{k}}}^{l}\right)$
with
\[
\begin{array}{l}
\left[\varGamma\right]_{\textbf{\textit{k}}}^{(1)}=+\frac{\Delta h}{U_{\textbf{\textit{k}}}^{2}-D_{\textbf{\textit{k}}}^{2}}-\frac{\Delta\left[U_{\textbf{\textit{k}}}^{2}+\left(h^{2}-\textbf{\textit{k}}^{2}\lambda^{2}-E_{\textbf{\textit{k}}}^{2}\right)\right]}{2U_{\textbf{\textit{k}}}\left(U_{\textbf{\textit{k}}}^{2}-D_{\textbf{\textit{k}}}^{2}\right)},\\
\left[\varGamma\right]_{\textbf{\textit{k}}}^{(-1)}=+\frac{\Delta h}{U_{\textbf{\textit{k}}}^{2}-D_{\textbf{\textit{k}}}^{2}}+\frac{\Delta\left[U_{\textbf{\textit{k}}}^{2}+\left(h^{2}-\textbf{\textit{k}}^{2}\lambda^{2}-E_{\textbf{\textit{k}}}^{2}\right)\right]}{2U_{\textbf{\textit{k}}}\left(U_{\textbf{\textit{k}}}^{2}-D_{\textbf{\textit{k}}}^{2}\right)},\\
\left[\varGamma\right]_{\textbf{\textit{k}}}^{(2)}=-\frac{\Delta h}{U_{\textbf{\textit{k}}}^{2}-D_{\textbf{\textit{k}}}^{2}}+\frac{\Delta\left[D_{\textbf{\textit{k}}}^{2}+\left(h^{2}-\textbf{\textit{k}}^{2}\lambda^{2}-E_{\textbf{\textit{k}}}^{2}\right)\right]}{2D_{\textbf{\textit{k}}}\left(U_{\textbf{\textit{k}}}^{2}-D_{\textbf{\textit{k}}}^{2}\right)},\\
\left[\varGamma\right]_{\textbf{\textit{k}}}^{(-2)}=-\frac{\Delta h}{U_{\textbf{\textit{k}}}^{2}-D_{\textbf{\textit{k}}}^{2}}-\frac{\Delta\left[D_{\textbf{\textit{k}}}^{2}+\left(h^{2}-\textbf{\textit{k}}^{2}\lambda^{2}-E_{\textbf{\textit{k}}}^{2}\right)\right]}{2D_{\textbf{\textit{k}}}\left(U_{\textbf{\textit{k}}}^{2}-D_{\textbf{\textit{k}}}^{2}\right)}.
\end{array}
\]

$S\left(\textbf{\textit{k}},\omega\right)=\sum_{l}\left[S\right]_{\textbf{\textit{k}}}^{l}/\left(\omega-E_{\textbf{\textit{k}}}^{l}\right)$ with

\[
\begin{array}{l}
\left[S\right]_{\textbf{\textit{k}}}^{(1)}=\left(k_{y}-ik_{x}\right)\lambda\left[+\frac{\xi_{\textbf{\textit{k}}}}{U_{\textbf{\textit{k}}}^{2}-D_{\textbf{\textit{k}}}^{2}}+\frac{U_{\textbf{\textit{k}}}^{2}+\left(\xi_{\textbf{\textit{k}}}^{2}-h^{2}-\textbf{\textit{k}}^{2}\lambda^{2}-\Delta^{2}\right)}{2U_{\textbf{\textit{k}}}\left(U_{\textbf{\textit{k}}}^{2}-D_{\textbf{\textit{k}}}^{2}\right)}\right],\\
\left[S\right]_{\textbf{\textit{k}}}^{(-1)}=\left(k_{y}-ik_{x}\right)\lambda\left[+\frac{\xi_{\textbf{\textit{k}}}}{U_{\textbf{\textit{k}}}^{2}-D_{\textbf{\textit{k}}}^{2}}-\frac{U_{\textbf{\textit{k}}}^{2}+\left(\xi_{\textbf{\textit{k}}}^{2}-h^{2}-\textbf{\textit{k}}^{2}\lambda^{2}-\Delta^{2}\right)}{2U_{\textbf{\textit{k}}}\left(U_{\textbf{\textit{k}}}^{2}-D_{\textbf{\textit{k}}}^{2}\right)}\right],\\
\left[S\right]_{\textbf{\textit{k}}}^{(2)}=\left(k_{y}-ik_{x}\right)\lambda\left[-\frac{\xi_{\textbf{\textit{k}}}}{U_{\textbf{\textit{k}}}^{2}-D_{\textbf{\textit{k}}}^{2}}-\frac{D_{\textbf{\textit{k}}}^{2}+\left(\xi_{\textbf{\textit{k}}}^{2}-h^{2}-\textbf{\textit{k}}^{2}\lambda^{2}-\Delta^{2}\right)}{2D_{\textbf{\textit{k}}}\left(U_{\textbf{\textit{k}}}^{2}-D_{\textbf{\textit{k}}}^{2}\right)}\right],\\
\left[S\right]_{\textbf{\textit{k}}}^{(-2)}=\left(k_{y}-ik_{x}\right)\lambda\left[-\frac{\xi_{\textbf{\textit{k}}}}{U_{\textbf{\textit{k}}}^{2}-D_{\textbf{\textit{k}}}^{2}}+\frac{D_{\textbf{\textit{k}}}^{2}+\left(\xi_{\textbf{\textit{k}}}^{2}-h^{2}-\textbf{\textit{k}}^{2}\lambda^{2}-\Delta^{2}\right)}{2D_{\textbf{\textit{k}}}\left(U_{\textbf{\textit{k}}}^{2}-D_{\textbf{\textit{k}}}^{2}\right)}\right].
\end{array}
\]
$F_{1}\left(\textbf{\textit{k}},\omega\right)=\sum_{l}\left[F_{1}\right]_{\textbf{\textit{k}}}^{l}/\left(\omega-E_{\textbf{\textit{k}}}^{l}\right)$
with
\[
\begin{array}{c}
\left[F_{1}\right]_{\textbf{\textit{k}}}^{(1)}=-\left[F_{1}\right]_{\textbf{\textit{k}}}^{(-1)}=+\frac{\Delta\lambda\left(k_{y}+ik_{x}\right)\left(\xi_{\textbf{\textit{k}}}+h\right)}{U_{\textbf{\textit{k}}}\left(U_{\textbf{\textit{k}}}^{2}-D_{\textbf{\textit{k}}}^{2}\right)},\\
\left[F_{1}\right]_{\textbf{\textit{k}}}^{(2)}=-\left[F_{1}\right]_{\textbf{\textit{k}}}^{(-2)}=-\frac{\Delta\lambda\left(k_{y}+ik_{x}\right)\left(\xi_{\textbf{\textit{k}}}+h\right)}{D_{\textbf{\textit{k}}}\left(U_{\textbf{\textit{k}}}^{2}-D_{\textbf{\textit{k}}}^{2}\right)}.
\end{array}
\]

$F_{2}\left(\textbf{\textit{k}},\omega\right)=\sum_{l}\left[F_{2}\right]_{\textbf{\textit{k}}}^{l}/\left(\omega-E_{\textbf{\textit{k}}}^{l}\right)$
with
\[
\begin{array}{c}
\left[F_{2}\right]_{\textbf{\textit{k}}}^{(1)}=-\left[F_{2}\right]_{\textbf{\textit{k}}}^{(-1)}=-\frac{\Delta\lambda\left(k_{y}-ik_{x}\right)\left(\xi_{\textbf{\textit{k}}}-h\right)}{U_{\textbf{\textit{k}}}\left(U_{\textbf{\textit{k}}}^{2}-D_{\textbf{\textit{k}}}^{2}\right)},\\
\left[F_{2}\right]_{\textbf{\textit{k}}}^{(2)}=-\left[F_{2}\right]_{\textbf{\textit{k}}}^{(-2)}=+\frac{\Delta\lambda\left(k_{y}-ik_{x}\right)\left(\xi_{\textbf{\textit{k}}}-h\right)}{D_{\textbf{\textit{k}}}\left(U_{\textbf{\textit{k}}}^{2}-D_{\textbf{\textit{k}}}^{2}\right)}.
\end{array}
\]

The expressions of all 9 independent matrix elements in mean-field
response function $A$ are

$A_{11}=+\underset{\textbf{\textit{p}}ll'}{\sum}\left[G_{1}\right]_{\textbf{\textit{p}}}^{l}\left[G_{1}\right]_{\textbf{\textit{p}}+\textbf{\textit{q}}}^{l'}\frac{f\left(E_{\textbf{\textit{p}}}^{l}\right)-f\left(E_{\textbf{\textit{p}}+\textbf{\textit{q}}}^{l'}\right)}{i\omega_{n}+E_{\textbf{\textit{p}}}^{l}-E_{\textbf{\textit{p}}+\textbf{\textit{q}}}^{l'}},$

$A_{12}=-\underset{\textbf{\textit{p}}ll'}{\sum}\left[\varGamma\right]_{\textbf{\textit{p}}}^{l}\left[\varGamma\right]_{\textbf{\textit{p}}+\textbf{\textit{q}}}^{l'}\frac{f\left(E_{\textbf{\textit{p}}}^{l}\right)-f\left(E_{\textbf{\textit{p}}+\textbf{\textit{q}}}^{l'}\right)}{i\omega_{n}+E_{\textbf{\textit{p}}}^{l}-E_{\textbf{\textit{p}}+\textbf{\textit{q}}}^{l'}},$

$A_{13}=+\underset{\textbf{\textit{p}}ll'}{\sum}\left[G_{1}\right]_{\textbf{\textit{p}}}^{l}\left[\varGamma\right]_{\textbf{\textit{p}}+\textbf{\textit{q}}}^{l'}\frac{f\left(E_{\textbf{\textit{p}}}^{l}\right)-f\left(E_{\textbf{\textit{p}}+\textbf{\textit{q}}}^{l'}\right)}{i\omega_{n}+E_{\textbf{\textit{p}}}^{l}-E_{\textbf{\textit{p}}+\textbf{\textit{q}}}^{l'}},$

$A_{14}=+\underset{\textbf{\textit{p}}ll'}{\sum}\left[\varGamma\right]_{\textbf{\textit{p}}}^{l}\left[G_{1}\right]_{\textbf{\textit{p}}+\textbf{\textit{q}}}^{l'}\frac{f\left(E_{\textbf{\textit{p}}}^{l}\right)-f\left(E_{\textbf{\textit{p}}+\textbf{\textit{q}}}^{l'}\right)}{i\omega_{n}+E_{\textbf{\textit{p}}}^{l}-E_{\textbf{\textit{p}}+\textbf{\textit{q}}}^{l'}},$

$A_{22}=+\underset{\textbf{\textit{p}}ll'}{\sum}\left[G_{2}\right]_{\textbf{\textit{p}}}^{l}\left[G_{2}\right]_{\textbf{\textit{p}}+\textbf{\textit{q}}}^{l'}\frac{f\left(E_{\textbf{\textit{p}}}^{l}\right)-f\left(E_{\textbf{\textit{p}}+\textbf{\textit{q}}}^{l'}\right)}{i\omega_{n}+E_{\textbf{\textit{p}}}^{l}-E_{\textbf{\textit{p}}+\textbf{\textit{q}}}^{l'}},$

$A_{23}=-\underset{\textbf{\textit{p}}ll'}{\sum}\left[G_{2}\right]_{\textbf{\textit{p}}}^{l}\left[\varGamma\right]_{\textbf{\textit{p}}+\textbf{\textit{q}}}^{-l'}\frac{f\left(E_{\textbf{\textit{p}}}^{l}\right)-f\left(E_{\textbf{\textit{p}}+\textbf{\textit{q}}}^{l'}\right)}{i\omega_{n}+E_{\textbf{\textit{p}}}^{l}-E_{\textbf{\textit{p}}+\textbf{\textit{q}}}^{l'}},$

$A_{24}=-\underset{\textbf{\textit{p}}ll'}{\sum}\left[\varGamma\right]_{\textbf{\textit{p}}}^{-l}\left[G_{2}\right]_{\textbf{\textit{p}}+\textbf{\textit{q}}}^{l'}\frac{f\left(E_{\textbf{\textit{p}}}^{l}\right)-f\left(E_{\textbf{\textit{p}}+\textbf{\textit{q}}}^{l'}\right)}{i\omega_{n}+E_{\textbf{\textit{p}}}^{l}-E_{\textbf{\textit{p}}+\textbf{\textit{q}}}^{l'}},$

$A_{34}=+\underset{\textbf{\textit{p}}ll'}{\sum}\left[G_{2}\right]_{\textbf{\textit{p}}}^{-l}\left[G_{1}\right]_{\textbf{\textit{p}}+\textbf{\textit{q}}}^{l'}\frac{f\left(E_{\textbf{\textit{p}}}^{l}\right)-f\left(E_{\textbf{\textit{p}}+\textbf{\textit{q}}}^{l'}\right)}{i\omega_{n}+E_{\textbf{\textit{p}}}^{l}-E_{\textbf{\textit{p}}+\textbf{\textit{q}}}^{l'}},$

$A_{43}=+\underset{\textbf{\textit{p}}ll'}{\sum}\left[G_{1}\right]_{\textbf{\textit{p}}}^{l}\left[G_{2}\right]_{\textbf{\textit{p}}+\textbf{\textit{q}}}^{-l'}\frac{f\left(E_{\textbf{\textit{p}}}^{l}\right)-f\left(E_{\textbf{\textit{p}}+\textbf{\textit{q}}}^{l'}\right)}{i\omega_{n}+E_{\textbf{\textit{p}}}^{l}-E_{\textbf{\textit{p}}+\textbf{\textit{q}}}^{l'}},$
where $f\left(x\right)=1/\left(e^{x/T}+1\right)$ is the Fermi-Dirac
distribution function. The expressions of 16 independent matrix elements
in mean-field response function $B$ are

$B_{11}=-\underset{\textbf{\textit{p}}ll'}{\sum}\left[F_{1}^{*}\right]_{\textbf{\textit{p}}}^{l}\left[F_{1}\right]_{\textbf{\textit{p}}+\textbf{\textit{q}}}^{l'}\frac{f\left(E_{\textbf{\textit{p}}}^{l}\right)-f\left(E_{\textbf{\textit{p}}+\textbf{\textit{q}}}^{l'}\right)}{i\omega_{n}+E_{\textbf{\textit{p}}}^{l}-E_{\textbf{\textit{p}}+\textbf{\textit{q}}}^{l'}},$

$B_{12}=+\underset{\textbf{\textit{p}}ll'}{\sum}\left[S\right]_{\textbf{\textit{p}}}^{l}\left[S^{*}\right]_{\textbf{\textit{p}}+\textbf{\textit{q}}}^{l'}\frac{f\left(E_{\textbf{\textit{p}}}^{l}\right)-f\left(E_{\textbf{\textit{p}}+\textbf{\textit{q}}}^{l'}\right)}{i\omega_{n}+E_{\textbf{\textit{p}}}^{l}-E_{\textbf{\textit{p}}+\textbf{\textit{q}}}^{l'}},$

$B_{13}=-\underset{\textbf{\textit{p}}ll'}{\sum}\left[S\right]_{\textbf{\textit{p}}}^{l}\left[F_{1}\right]_{\textbf{\textit{p}}+\textbf{\textit{q}}}^{l'}\frac{f\left(E_{\textbf{\textit{p}}}^{l}\right)-f\left(E_{\textbf{\textit{p}}+\textbf{\textit{q}}}^{l'}\right)}{i\omega_{n}+E_{\textbf{\textit{p}}}^{l}-E_{\textbf{\textit{p}}+\textbf{\textit{q}}}^{l'}},$

$B_{14}=-\underset{\textbf{\textit{p}}ll'}{\sum}\left[F_{1}^{*}\right]_{\textbf{\textit{p}}}^{l}\left[S^{*}\right]_{\textbf{\textit{p}}+\textbf{\textit{q}}}^{l'}\frac{f\left(E_{\textbf{\textit{p}}}^{l}\right)-f\left(E_{\textbf{\textit{p}}+\textbf{\textit{q}}}^{l'}\right)}{i\omega_{n}+E_{\textbf{\textit{p}}}^{l}-E_{\textbf{\textit{p}}+\textbf{\textit{q}}}^{l'}},$

$B_{21}=+\underset{\textbf{\textit{p}}ll'}{\sum}\left[S^{*}\right]_{\textbf{\textit{p}}}^{l}\left[S\right]_{\textbf{\textit{p}}+\textbf{\textit{q}}}^{l'}\frac{f\left(E_{\textbf{\textit{p}}}^{l}\right)-f\left(E_{\textbf{\textit{p}}+\textbf{\textit{q}}}^{l'}\right)}{i\omega_{n}+E_{\textbf{\textit{p}}}^{l}-E_{\textbf{\textit{p}}+\textbf{\textit{q}}}^{l'}},$

$B_{22}=-\underset{\textbf{\textit{p}}ll'}{\sum}\left[F_{2}\right]_{\textbf{\textit{p}}}^{l}\left[F_{2}^{*}\right]_{\textbf{\textit{p}}+\textbf{\textit{q}}}^{l'}\frac{f\left(E_{\textbf{\textit{p}}}^{l}\right)-f\left(E_{\textbf{\textit{p}}+\textbf{\textit{q}}}^{l'}\right)}{i\omega_{n}+E_{\textbf{\textit{p}}}^{l}-E_{\textbf{\textit{p}}+\textbf{\textit{q}}}^{l'}},$

$B_{23}=+\underset{\textbf{\textit{p}}ll'}{\sum}\left[S^{*}\right]_{\textbf{\textit{p}}}^{l}\left[F_{2}\right]_{\textbf{\textit{p}}+\textbf{\textit{q}}}^{l'}\frac{f\left(E_{\textbf{\textit{p}}}^{l}\right)-f\left(E_{\textbf{\textit{p}}+\textbf{\textit{q}}}^{l'}\right)}{i\omega_{n}+E_{\textbf{\textit{p}}}^{l}-E_{\textbf{\textit{p}}+\textbf{\textit{q}}}^{l'}},$

$B_{24}=+\underset{\textbf{\textit{p}}ll'}{\sum}\left[F_{2}^{*}\right]_{\textbf{\textit{p}}}^{l}\left[S\right]_{\textbf{\textit{p}}+\textbf{\textit{q}}}^{l'}\frac{f\left(E_{\textbf{\textit{p}}}^{l}\right)-f\left(E_{\textbf{\textit{p}}+\textbf{\textit{q}}}^{l'}\right)}{i\omega_{n}+E_{\textbf{\textit{p}}}^{l}-E_{\textbf{\textit{p}}+\textbf{\textit{q}}}^{l'}},$

$B_{31}=-\underset{\textbf{\textit{p}}ll'}{\sum}\left[F_{1}\right]_{\textbf{\textit{p}}}^{l}\left[S\right]_{\textbf{\textit{p}}+\textbf{\textit{q}}}^{l'}\frac{f\left(E_{\textbf{\textit{p}}}^{l}\right)-f\left(E_{\textbf{\textit{p}}+\textbf{\textit{q}}}^{l'}\right)}{i\omega_{n}+E_{\textbf{\textit{p}}}^{l}-E_{\textbf{\textit{p}}+\textbf{\textit{q}}}^{l'}},$

$B_{32}=+\underset{\textbf{\textit{p}}ll'}{\sum}\left[F_{2}\right]_{\textbf{\textit{p}}}^{l}\left[S^{*}\right]_{\textbf{\textit{p}}+\textbf{\textit{q}}}^{l'}\frac{f\left(E_{\textbf{\textit{p}}}^{l}\right)-f\left(E_{\textbf{\textit{p}}+\textbf{\textit{q}}}^{l'}\right)}{i\omega_{n}+E_{\textbf{\textit{p}}}^{l}-E_{\textbf{\textit{p}}+\textbf{\textit{q}}}^{l'}},$

$B_{33}=-\underset{\textbf{\textit{p}}ll'}{\sum}\left[F_{2}\right]_{\textbf{\textit{p}}}^{l}\left[F_{1}\right]_{\textbf{\textit{p}}+\textbf{\textit{q}}}^{l'}\frac{f\left(E_{\textbf{\textit{p}}}^{l}\right)-f\left(E_{\textbf{\textit{p}}+\textbf{\textit{q}}}^{l'}\right)}{i\omega_{n}+E_{\textbf{\textit{p}}}^{l}-E_{\textbf{\textit{p}}+\textbf{\textit{q}}}^{l'}},$

$B_{34}=+\underset{\textbf{\textit{p}}ll'}{\sum}\left[S\right]_{\textbf{\textit{p}}}^{-l}\left[S^{*}\right]_{\textbf{\textit{p}}+\textbf{\textit{q}}}^{l'}\frac{f\left(E_{\textbf{\textit{p}}}^{l}\right)-f\left(E_{\textbf{\textit{p}}+\textbf{\textit{q}}}^{l'}\right)}{i\omega_{n}+E_{\textbf{\textit{p}}}^{l}-E_{\textbf{\textit{p}}+\textbf{\textit{q}}}^{l'}},$

$B_{41}=-\underset{\textbf{\textit{p}}ll'}{\sum}\left[S^{*}\right]_{\textbf{\textit{p}}}^{l}\left[F_{1}^{*}\right]_{\textbf{\textit{p}}+\textbf{\textit{q}}}^{l'}\frac{f\left(E_{\textbf{\textit{p}}}^{l}\right)-f\left(E_{\textbf{\textit{p}}+\textbf{\textit{q}}}^{l'}\right)}{i\omega_{n}+E_{\textbf{\textit{p}}}^{l}-E_{\textbf{\textit{p}}+\textbf{\textit{q}}}^{l'}}.$

$B_{42}=+\underset{\textbf{\textit{p}}ll'}{\sum}\left[S\right]_{\textbf{\textit{p}}}^{l}\left[F_{2}^{*}\right]_{\textbf{\textit{p}}+\textbf{\textit{q}}}^{l'}\frac{f\left(E_{\textbf{\textit{p}}}^{l}\right)-f\left(E_{\textbf{\textit{p}}+\textbf{\textit{q}}}^{l'}\right)}{i\omega_{n}+E_{\textbf{\textit{p}}}^{l}-E_{\textbf{\textit{p}}+\textbf{\textit{q}}}^{l'}}.$

$B_{43}=+\underset{\textbf{\textit{p}}ll'}{\sum}\left[S\right]_{\textbf{\textit{p}}}^{l}\left[S^{*}\right]_{\textbf{\textit{p}}+\textbf{\textit{q}}}^{-l'}\frac{f\left(E_{\textbf{\textit{p}}}^{l}\right)-f\left(E_{\textbf{\textit{p}}+\textbf{\textit{q}}}^{l'}\right)}{i\omega_{n}+E_{\textbf{\textit{p}}}^{l}-E_{\textbf{\textit{p}}+\textbf{\textit{q}}}^{l'}}.$

$B_{44}=-\underset{\textbf{\textit{p}}ll'}{\sum}\left[F_{1}^{*}\right]_{\textbf{\textit{p}}}^{l}\left[F_{2}^{*}\right]_{\textbf{\textit{p}}+\textbf{\textit{q}}}^{l'}\frac{f\left(E_{\textbf{\textit{p}}}^{l}\right)-f\left(E_{\textbf{\textit{p}}+\textbf{\textit{q}}}^{l'}\right)}{i\omega_{n}+E_{\textbf{\textit{p}}}^{l}-E_{\textbf{\textit{p}}+\textbf{\textit{q}}}^{l'}}.$

\end{document}